\documentclass[iop,apj,twocolumn]{emulateapj}
\usepackage{times}
\usepackage{epsfig}
\usepackage{amsmath, amsthm, amssymb}
\usepackage[plainpages=false, colorlinks=true, anchorcolor=blue, linkcolor=blue, citecolor=blue, bookmarks=false]
{hyperref}
\usepackage{color}
\usepackage{microtype}
\usepackage{float}
\usepackage[caption = false]{subfig}
\usepackage{graphicx}

\newcommand{\beq}{\begin{equation}}
\newcommand{\eeq}{\end{equation}}
\newcommand{\bea}{\begin{eqnarray}}
\newcommand{\eea}{\end{eqnarray}}

\slugcomment{Draft version - \today}

\begin{document}

\title{Radiation Transport for Explosive Outflows: A Multigroup Hybrid Monte Carlo Method} 
\author{Ryan T. Wollaeger$^{1}$, Daniel R. van~Rossum$^{2}$, Carlo Graziani$^{2}$, Sean M. Couch$^{2,3}$,
\\George C. Jordan IV$^{2}$, Donald Q. Lamb$^{2}$, and Gregory A. Moses$^{1}$}

\affil{$^{1}$Department of Nuclear Engineering \& Engineering Physics, University of Wisconsin, Madison
1500 Engineering Drive, 410 ERB, Madison, WI, 53706; wollaeger@wisc.edu}
\affil{$^{2}$Flash Center for Computational Science, Department of Astronomy
  \& Astrophysics, University of Chicago, Chicago,
IL, 
60637; flash.uchicago.edu}
\affil{$^{3}$Hubble Fellow}

\shorttitle{IMC-DDMC}

\begin{abstract}

We explore Implicit Monte Carlo (IMC) and Discrete Diffusion Monte Carlo (DDMC)
for radiation transport in high-velocity outflows with structured opacity.
The IMC method is a stochastic computational
technique for nonlinear radiation transport.  IMC is partially implicit in time and may suffer in
efficiency when tracking Monte Carlo particles through optically thick materials.  DDMC
accelerates IMC in diffusive domains.  Abdikamalov
extended IMC and DDMC to multigroup, velocity-dependent transport with the intent of modeling
neutrino dynamics in core-collapse supernovae.  Densmore has also formulated a multifrequency extension
to the originally grey DDMC method.
We rigorously formulate IMC and DDMC over a high-velocity Lagrangian
grid for possible application to photon transport in the post-explosion phase of Type Ia supernovae.
This formulation includes an analysis that yields an additional factor in the
standard IMC-to-DDMC spatial interface condition.  To our knowledge the new boundary condition is distinct
from others presented in prior DDMC literature.
The method is suitable for a variety of opacity distributions and may be
applied to semi-relativistic radiation transport in simple fluids and geometries.
Additionally, we test the code, called {\tt SuperNu}, using an analytic solution
having static material, as well as with a
manufactured solution for moving material with structured opacities.  Finally, we
demonstrate with a simple source and 10 group logarithmic wavelength grid that IMC-DDMC performs
better than pure IMC in terms of accuracy and speed when there are large disparities between the magnitudes
of opacities in adjacent groups.  We also present and test our implementation
of the new boundary condition.


\end{abstract}

\section{Introduction}
\label{sec:Intro}

Type Ia supernovae (SNe Ia) are thermonuclear explosions
of carbon-oxygen white dwarf stars.  A variety of models
have been proposed for SNe Ia.  In all of these models, the 
expansion becomes ballistic and homologous within $\sim$100 
seconds.  Gamma rays from the radioactive decay of 
$^{56}$Ni formed in the explosion heat the expanding ejecta, 
making it glow for weeks.  Two remarkable features of Type
Ia supernova (SN Ia) light curves are that their peak luminosities 
span a modest range and that they can be calibrated to be 
standard candles, making them useful for measuring 
distances in the universe~[see, e.g.,~\citet{riess1998,perlmutter1999}].

SN Ia light curves and spectra are the result of complex 
radiative processes involving the interaction of photons with
millions of spectral lines in various stages of ionization, 
including strong scattering effects, in the asymmetric,
chemically inhomogeneous, quasi-relativistically expanding
ejecta~[see, e.g.,~\citet{kasen2006,baron2007,vanrossum2012}].

Nevertheless, if provided with quality material data,
a robust numerical
radiation transport method that is amenable to 
parallelization and is efficient in optically thick
regions of phase space should be able to
reproduce SN Ia spectra and light curves.  Monte Carlo (MC) is a
simple approach to transport computation that allows one
to model ``bundles'' of photons directly as particle
histories.  Particle processes are carried out stochastically
with random number sampling~\citep[p.~296]{lewis1993}.  Each of
these histories is independent from the others; hence Monte Carlo
can be performed in parallel and in domain decomposed 
settings.


%

Implicit Monte Carlo (IMC) is an MC
technique for solving the time dependent radiation
transport equation coupled nonlinearly with material 
\citep{fleck1971}.  In implementation, a non-dimensional
quantity resulting from temporal discretization of the
material equation dictates how likely a Monte Carlo
particle can be absorbed or re-emitted instantaneously.
The non-dimensional quantity is referred to as the
Fleck factor.\footnote{Note the Fleck factor is not a
directly tunable parameter but follows naturally from
linearizing the thermal transport equations within each
time step.}  The instantaneous absorption-reemission
event in an IMC time step can be modeled as an effective
inelastic scattering event for each particle history.

The semi-implicitness of IMC provides an advantage over
explicit radiation transport methods by mitigating
Courant-type instabilities due to large time steps and optically
thick domains~\citep{fleck1971}.  This advantage is obtained
through the isotropic effective scattering events mentioned
above.  Explicit transport methods incur these errors from
having to instantiate the entire emission energy as particles
at the beginning of the time step.  With IMC, a large time 
step and a strong absorption opacity allow effective scattering 
to dominate MC particle-material grid interactions.

A significant drawback of IMC is the computational inefficiency
of having to model instantaneous remission through effective
scattering in optically thick regimes.  Both effective and physical
scattering can be partly avoided by hybridizing IMC with a 
diffusion method.  This diffusion routine can either be 
deterministic or stochastic.  \cite{fleck1984} 
incorporate Random Walk (RW) to replace small scattering 
steps with large diffusion steps.  These large diffusion steps
assume a particle has undergone several collisions and may
be isotropically placed on a sphere centered at the particle's
initial position.  The diffusion sphere radius is bounded by
the cell in which the particle resides.  As Densmore and others note 
\citep{densmore2007,densmore2012}, the RW method must use
transport for particles near spatial cell boundaries even in
optically thick domains; so its ability to increase IMC
efficiency is limited.

Discrete Diffusion Monte Carlo (DDMC)~\citep{densmore2007,densmore2008,densmore2012}
and Implicit Monte Carlo Diffusion (IMD)~\citep{gentile2001,cleveland2010}
are recent alternatives to Random Walk that lend a stochastic
interpretation to the discretized diffusion equation.
Consequently in either method, a diffusion Monte Carlo particle's
position is fundamentally ambiguous within the spatial cell where
it resides.  The core difference between IMD and DDMC is
the treatment of particle histories.  In IMD, the
diffusion equation is discrete in time and there is a probability
to determine completion of each time step~\citep{cleveland2010}.
DDMC treats particle times continuously, removing
causal ambiguity when interfaced or hybridized with IMC
\citep{densmore2007}.

Multifrequency Implicit Monte Carlo-Discrete Diffusion Monte 
Carlo (IMC-DDMC) methods have very recently been formulated
by~\cite{densmore2012} and~\cite{abdikamalov2012}.  Densmore's 
formulation is for local thermodynamic equilibrium (LTE)
photon transport with an opacity dependence on frequency that
is roughly monotonic.  Specifically, it is assumed that the
opacity is optically thick at low photon frequency and can be
modeled with grey DDMC while multifrequency IMC is applied
above a user-defined frequency threshold.  This frequency
threshold depends heuristically on the cell-local material
properties.  In practice, this cutoff can be achieved
approximately on a group structure by lumping adjacent groups
that are sufficiently diffuse into one large DDMC group.

The formulation of~\cite{abdikamalov2012} is for neutrino
transport in either static or velocity-dependent material.
The frequency effects are treated with a multigroup approach and,
similarly to the approach of~\cite{densmore2012}, a heuristic
determines whether IMC or DDMC is applied at a specific spatial
cell and frequency group.

Here, we present an extension of IMC-DDMC to photon transport in SNe Ia that is similar to that of
\cite{abdikamalov2012} for neutrino transport in core collapse
supernovae.  In contrast to the work of~\cite{abdikamalov2012},
we treat velocity as linearly continuous at the sub-cell level; we
test an IMC-DDMC-specific algorithm to account for spatial grid
motion; we only apply first-order relativity to both IMC and DDMC;
we derive a new IMC-DDMC boundary condition for semi-relativistic outflow;
and we describe methods for non-uniform group structuring
that extend formulae presented by~\cite{densmore2012}.
We also make note of the consequence of DDMC particle frequency
ambiguity in Doppler shifting over multigroup structures.
As done by~\cite{kasen2006}, we exploit the
homologous relation between space and fluid expansion time and
formulate the method over a velocity grid.  Additionally, we use
the Method of Manufactured Solutions~\citep[p.~219]{oberkampf2010} to verify
{\tt SuperNu}'s ability to reproduce radiation energy density profiles with
appropriate sources and initial conditions.

To our knowledge, the modification to the standard IMC-DDMC boundary
condition described in Section~\ref{subsec:MBLA} is a novel theoretical
finding.  Specifically, we obtain an additional term that
multiplies the probability an IMC particle incident on a
DDMC region will convert to DDMC.  This factor is singular for incident
particles that have comoving directions lying in the tangent plane of the IMC-DDMC
boundary at the particle's point of contact.  However, we show the expected IMC
energy current reflected and transmitted from and into the DDMC region is finite.
Since the singularity does not introduce infinities in energy balance, we
suppose that the new quantity may be interpreted as an IMC particle weight
modification.

This paper is organized as follows.  In Section~\ref{sec:IMCDDMC},
we discuss the IMC-DDMC theory and implementation.  We
first discuss the treatment of fluid coupling and relativistic effects; in
Section~\ref{subsec:LFIMC}, we review and formulate IMC for lab frame transport
on a velocity grid; in Section~\ref{subsec:CFDDMC}, we review and discuss
DDMC; in Section~\ref{subsec:MBLA}, we then perform an asymptotic
analysis on a moving surface to obtain the new boundary condition;
in Section~\ref{subsec:MFIMCDDMC}, we move the discussion to coupling IMC
and DDMC over a high-velocity Lagrangian grid.  In Section~\ref{sec:Static}, we
exhibit a closed form verification test for static material radiation transport.
This test is an extension of the thermally coupled P$_{1}$ solutions
provided by~\cite{mcclarren2008} to account for rudimentary multifrequency.
In Section~\ref{sec:Manu}, we test our implementation of IMC-DDMC
against a manufactured solution that includes outflow and a multigroup
structure.  The manufactured solution is constructed to counteract some
recognized properties of radiation trapped in high velocity, spherical
flow such as inverse quartic dependence of energy density on the Eulerian radius
\citep[p.~474]{mihalas1984}. In Section~\ref{sec:Heav}, we test the efficiency
of IMC-DDMC relative to pure IMC for group structures of varying
contrast ratios between alternating thick-thin grouped opacity values.
Also in Section~\ref{sec:Heav}, we demonstrate a discrepancy between
IMC-DDMC and pure IMC may form at IMC-DDMC interfaces for high velocity outflows.
We find this error to be large for IMC-DDMC thresholds on the order of 10 mean
free paths per cell per group in the set of Heaviside source problems discussed
(in other words, when IMC is applied in a cell and group with fewer than 10 mean
free paths across some cell length measure and DDMC is applied otherwise).
We implement the new boundary condition in IMC-DDMC
and test its ability to counteract the redshift-induced error for coupling
thresholds equal to 3 mean free paths and to 10 mean free paths per cell per
group.
Finally in Section~\ref{sec:Conc}, we summarize our findings,
discuss possible future work, and discuss the feasibility of the general
multifrequency treatment of IMC-DDMC.

\section{Homologous Velocity Space IMC-DDMC}
\label{sec:IMCDDMC}

The derivation of IMC-DDMC and IMC-IMD in a multigroup setting has been
discussed extensively in prior publications
\citep{cleveland2010,densmore2012,abdikamalov2012}.  We briefly
review the relevant method derivations and discuss the distinctive
algorithmic features we have implemented to reconcile IMC-DDMC 
with a ballistic fluid on a Lagrangian grid.  
Subsequently, we describe an optimization method we will refer
as ``group lumping'', Eqs.~\eqref{eq43}-\eqref{eq45}.  Group
lumping combines adjacent groups that are heuristically deemed
DDMC-appropriate into larger groups.  Hence the
method is a natural extension to methods that model all radiation below
a frequency threshold with grey DDMC~\citep{densmore2012}.
It is evident from theory that group lumping over optically-thick regions
in phase space will increase the overall code efficiency.

As Pomraning and Castor have done, we denote comoving quantities
with a subscript 0 and leave the corresponding lab (or outflow center)
frame quantities unsubscripted.
The thermal
radiative transport equation without external sources in the lab
frame is~\citep{pomraning1973,castor2004}
\begin{multline}
\label{eq1}
\frac{1}{c}\frac{\partial I_{\nu}}{\partial t}+\hat{\Omega}\cdot
\nabla I_{\nu}+\sigma_{\nu,a}I_{\nu}=\sigma_{\nu,a}B_{\nu}-
\sigma_{\nu,s}I_{\nu}+\\\int_{4\pi}\int_{0}^{\infty}\frac{\nu}{\nu'}
\sigma_{s}(\vec{r},\nu'\rightarrow\nu,\hat{\Omega}'\rightarrow
\hat{\Omega})I_{\nu'}(\vec{r},\hat{\Omega}',t)d\nu'd\Omega'
\end{multline}
where $\vec{r}$ is the spatial coordinate, $\hat{\Omega}$ is
a unit direction, $t$ is time, $\nu$ is frequency, $c$ is the speed
of light, $I_{\nu}$ is the radiation intensity, and $B_{\nu}$ is the 
lab frame thermal emission.  If the fluid is static, 
$B_{\nu}(\hat{\Omega})=B_{0,\nu_{0}}$ is the Planck function
\citep[p.~156]{pomraning1973}.  Anisotropies in the radiation field due
to fluid motion are represented in the opacities and
intensities in Eq.~\eqref{eq1}.

For a radiative hydrodynamic system, the Euler equations that couple
with Eq.~\eqref{eq1} are~\citep[p.~85]{castor2004}
\begin{equation}
\label{eq2}
\frac{\partial\rho}{\partial t}+\nabla\cdot(\rho\vec{U})=0 \;\;,
\end{equation}
\begin{equation}
\label{eq3}
\frac{\partial(\rho\vec{U})}{\partial t}+
\nabla\cdot(\rho\vec{U}\vec{U})+\nabla P = -\vec{g} \;\;,
\end{equation}
and
\begin{equation}
\label{eq4}
\frac{\partial}{\partial t}\left(\rho e+\frac{1}{2}\rho U^{2}\right)
+\nabla\cdot\left(\rho\vec{U}(e+P/\rho)+\frac{1}{2}\rho
\vec{U}U^{2}\right)=-g^{(0)}
\end{equation}
where $e$, $\rho$, $\vec{U}$, and $P$ are the internal energy, density,
velocity, and pressure of the fluid in the lab frame (an inertial frame
not following any particular fluid parcel).  The 4-vector $(g^{(0)},\vec{g})$
is the radiation energy-momentum coupling~\citep[p.~109]{castor2004}.  The superscript,
(0), denotes the time component of the 4-vector.  We are
interested in tracking changes to the material in the comoving or Lagrangian
coordinate of velocity cells.  The Lagrangian equation corresponding to
Eq.~\eqref{eq4} is~\citep[pp.~5-10]{castor2004}
\begin{equation}
\label{eq5}
\rho\frac{De}{Dt}+P\nabla\cdot\vec{U}=-g^{(0)}
\end{equation}
where the operator $D/Dt$ is the Lagrangian derivative. 
The second term on the left-hand-side of equation
\eqref{eq5} is usually negligible relative to $g^{(0)}$
in the physical regimes of interest.
In a homologous outflow, $\nabla\cdot\vec{U}=3/t_{\text{exp}}$ where $t_{\text{exp}}$
is the fluid expansion time.  The ratio of the rate of adiabatic, ideal gas
cooling to the thermal radiation deposition rate is approximately
$\Lambda=\displaystyle\frac{3N_{A}k}{M_{A}aT^{3}}\frac{\rho/c\sigma_{P}}{t_{\text{exp}}}$
where $N_{A}$, $k$, $M_{A}$, $T$, and $\sigma_{P}$ are
Avogadro's number, Boltzmann's constant, molar mass, temperature, and Planck
opacity, respectively~\citep{kasen2006}.  For an outflow of Nickel with $T=12,000$ K,
$\sigma_{P}=0.1\rho$ cm$^{-1}$, and $t_{\text{exp}}=10$ days, the ratio is
approximately $\Lambda\approx1.3\times 10^{-7}$.  As~\cite{kasen2006} argue, $\Lambda$ remains small over
the times of interest in light curve and spectra observation for Type Ia SNe.  Our 
outflow simulations employ numbers that generate small $\Lambda$.
So we neglect the adiabatic cooling rate of the gas, $P\nabla\cdot\vec{U}$, in our subsequent analysis.
We do not extend this approximation to photons; the adiabatic cooling in the radiation field is a
large factor in our velocity-dependent simulations.
The analysis provided agrees with arguments by~\citet{pinto2000} as well.

The Lagrangian momentum equation is~\citep[p.~9]{castor2004}
\begin{equation}
\label{eq6}
\rho\frac{D\vec{U}}{Dt}+\nabla P = -\vec{g} \;\;.
\end{equation}
The velocity across a cell (or discrete fluid parcel) will not
change in the Lagrangian frame; hence the first term of Eq.~\eqref{eq6}
on the left-hand-side will be zero in the homologous outflow.  Additionally, we
assume the pressure gradient across the fluid parcel will be small.
Consequently $\vec{g}=0$.  So $g_{0}^{(0)}= g^{(0)}-\vec{U}\cdot
\vec{g}/c^{2}\approx g^{(0)}$.  With the above simplifications, 
Eq.~\eqref{eq5} becomes~\citep{szoke2005}
\begin{multline}
\label{eq7}
C_{v}\frac{DT}{Dt}=-g_{0}^{(0)}=-g_{0,a}^{(0)}-g_{0,s}^{(0)}=\\
\int_{4\pi}\int_{0}^{\infty}\sigma_{0,\nu_{0},a}
(I_{0,\nu_{0}}-B_{0,\nu_{0}})d\nu_{0}d\Omega_{0}\\
+\int_{4\pi}\int_{0}^{\infty}\sigma_{0,\nu_{0},s}
I_{0,\nu_{0}}d\nu_{0}d\Omega_{0}-
\int_{4\pi}\int_{0}^{\infty}\int_{4\pi}\int_{0}^{\infty}\ldots\\
\frac{\nu_{0}}{\nu_{0}'}\sigma_{0,s}
(\vec{r},\nu_{0}'\rightarrow\nu_{0},\hat{\Omega}_{0}'\cdot\hat{\Omega}_{0})
I_{0,\nu_{0}'}d\nu_{0}'d\Omega_{0}'d\nu_{0}d\Omega_{0}
\;\;,
\end{multline}
where $C_{v}$ is the heat capacity per unit volume, $g_{0,a}^{(0)}$
($g_{0,s}^{(0)}$) includes all absorption (scattering) terms, and
\begin{equation}
\label{eq8}
\sigma_{0,\nu_{0},s}=\int_{4\pi}\int_{0}^{\infty}\sigma_{0,s}
(\vec{r},\nu_{0}\rightarrow\nu_{0}',
\hat{\Omega}_{0}\cdot\hat{\Omega}_{0}')d\nu_{0}'d\Omega_{0}' \;\;.
\end{equation}
Equation~\eqref{eq7} is amenable to the usual IMC temporal discretization
\citep{fleck1971,abdikamalov2012}.  The coupling, however, is in the 
comoving frame and Eq.~\eqref{eq1} is in the lab frame.  If the fluid
field is nowhere accelerating, the comoving transport equation to first
order is~\citep[p.~110]{castor2004}
\begin{multline}
\label{eq9}
(1+\hat{\Omega}_{0}\cdot\vec{U}/c)\frac{1}{c}\frac{DI_{0,\nu_{0}}}{Dt}\\+
\hat{\Omega}_{0}\cdot\nabla I_{0,\nu_{0}}-\frac{\nu_{0}}{c}\hat{\Omega}_{0}
\cdot\nabla\vec{U}\cdot\nabla_{\nu_{0}\hat{\Omega}_{0}}I_{0,\nu_{0}}\\+
\frac{3}{c}\hat{\Omega}_{0}\cdot\nabla\vec{U}\cdot\hat{\Omega}_{0}
I_{0,\nu_{0}} = \sigma_{0,\nu_{0},a}(B_{0,\nu_{0}}-I_{0,\nu_{0}})
-\sigma_{0,\nu_{0},s}I_{0,\nu_{0}}\\+
\int_{4\pi}\int_{0}^{\infty}\frac{\nu_{0}}{\nu_{0}'}\sigma_{0,s}
(\vec{r},\nu_{0}'\rightarrow\nu_{0},
\hat{\Omega}_{0}'\cdot\hat{\Omega}_{0})I_{0,\nu_{0}'}d\nu_{0}'d\Omega_{0}'
\;\;,
\end{multline}
where in Castor's notation $\nabla_{\nu_{0}\hat{\Omega}_{0}}$ is the
comoving momentum derivative for photons.

\subsection{Lab Frame IMC}
\label{subsec:LFIMC}

Applying the IMC discretization to Eqs.~\eqref{eq7} and
\eqref{eq9} and expressing the re-balanced equations in differential
form~\citep{densmore2012} gives
\begin{multline}
\label{eq10}
C_{v}\frac{DT}{Dt}=\\f_{n}\left(\int_{4\pi}\int_{0}^{\infty}
\sigma_{0,\nu_{0},a,n}
I_{0,\nu_{0}}d\nu_{0}d\Omega_{0}-\sigma_{P,n}acT_{n}^{4}\right)
-g_{0,s}^{(0)}
\end{multline}
and
\begin{multline}
\label{eq11}
(1+\hat{\Omega}_{0}\cdot\vec{U}/c)\frac{1}{c}\frac{DI_{0,\nu_{0}}}{Dt}+
\hat{\Omega}_{0}\cdot\nabla I_{0,\nu_{0}}\\-\frac{\nu_{0}}{c}\hat{\Omega}_{0}
\cdot\nabla\vec{U}\cdot\nabla_{\nu_{0}\hat{\Omega}_{0}}I_{0,\nu_{0}}+
\frac{3}{c}\hat{\Omega}_{0}\cdot\nabla\vec{U}\cdot\hat{\Omega}_{0}
I_{0,\nu_{0}}\\+(\sigma_{0,\nu_{0},s,n}+\sigma_{0,\nu_{0},a,n})I_{0,\nu_{0}}=
\frac{f_{n}}{4\pi}\sigma_{0,\nu_{0},a,n}b_{0,\nu_{0},n}acT_{n}^{4}\\+
\frac{b_{0,\nu_{0},n}\sigma_{0,\nu_{0},a,n}}{4\pi\sigma_{P,n}}(1-f_{n})\int_{4\pi}
\int_{0}^{\infty}\sigma_{0,\nu_{0}',a,n}I_{0,\nu_{0}'}d\nu_{0}'d\Omega_{0}'\\+
\int_{4\pi}\int_{0}^{\infty}\frac{\nu_{0}}{\nu_{0}'}\sigma_{0,s,n}
(\vec{r},\nu_{0}'\rightarrow\nu_{0},
\hat{\Omega}_{0}'\cdot\hat{\Omega}_{0})I_{0,\nu_{0}'}d\nu_{0}'d\Omega_{0}'
\;\;,
\end{multline}
where the integer subscript $n$ denotes quantities evaluated at the
beginning of a time step.  The value $g_{0,s}^{(0)}$ has been lumped entirely
into the material equation since physical scattering generally admits direct
treatment in MC transport.
$P$,$R$ and $g$ subscripts indicate Planck, Rosseland or grouped quantities,
respectively.  The opacities are evaluated in the comoving frame. The
value $b_{0,\nu_{0}}$ is the frequency-normalized Planck function in the comoving
frame, and the Fleck factor $f_{n}$~\citep{fleck1971} is
\begin{equation}
\label{eq12}
f_{n}=\frac{1}{1+\alpha\beta_{n}\sigma_{P,n}c\Delta t_{n}} \;\;.
\end{equation}
The value $\alpha\in[0,1]$ is a time centering control parameter (often set to 1),
$\beta_{n}=4aT_{n}^{3}/C_{v,n}$ and $\Delta t_{n}$ is the physical time step size for
time step $n$.  The second term on the right-hand-side of Eq.~\eqref{eq11} is the
source due to effective scattering~\citep{fleck1971,densmore2012}.  The differential
effective scattering opacity, $(1-f_{n})b_{0,\nu_{0},n}\sigma_{0,\nu_{0},a,n}
\sigma_{0,\nu_{0}',a,n}\nu_{0}'/4\pi\sigma_{P,n}\nu_{0}$, is separable in $\nu_{0}$ and
$\nu_{0}'$; so the new frequency of a photon undergoing effective scattering is
probabilistically independent of the old frequency~\cite[p.~327]{mihalas1984}.

Equation~\eqref{eq11} could in principle be replaced with the fully relativistic
comoving transport equation described by~\cite[p.~434]{mihalas1984}.
Here, we consider only first order relativistic effects but note that higher order
effects could be incorporated with relative ease in MC codes~\citep{abdikamalov2012}.
The typical outflow speed of Type Ia SNe is approximately $U_{\max}=30,000$
km/s, so $(U_{\max}/c)^{2}\approx0.01$.  The lab frame Eulerian coordinate is related
to the velocity through
\begin{equation}
\label{eq13}
\vec{r}=\vec{U}(t_{n}+t_{\min}) \;\;,
\end{equation}
where $t_{n}=\sum_{n'=1}^{n-1}\Delta t_{n}$ and $t_{\min}$ is the starting time of
the expansion.

The right hand side of Eq.~\eqref{eq11} implies source MC particles may be 
generated, scattered, or absorbed in the comoving frame in the same manner
as the static material IMC method~\citep{abdikamalov2012,lucy2005}.  To seed
interaction locations between events in a particle's history, the particle's
properties may be mapped to the lab frame and streamed by Eq.~\eqref{eq1}.
Neglecting all terms of order $U^{2}/c^{2}$ or higher, the frequency and
direction transformations are~\citep[p.~103]{castor2004}
\begin{equation}
\label{eq14}
\nu=\nu_{0}\left(1+\frac{\hat{\Omega}_{0}\cdot\vec{U}}{c}\right)
\end{equation}
and
\begin{equation}
\label{eq15}
\hat{\Omega}=\frac{\hat{\Omega}_{0}+\vec{U}/c}
{1+\hat{\Omega}_{0}\cdot\vec{U}/c}
\end{equation}
respectively.  The cumulative effect of these transformations in each particle
history accounts for Doppler shifting and aberration in the MC process
\citep{lucy2005}.  According to~\cite[p.~156]{pomraning1973} and~\cite[p.~104]{castor2004},
the opacity in the lab frame is
\begin{equation}
\label{eq16}
\sigma_{\text{lab}} = \frac{\nu_{0}}{\nu}\sigma_{\text{cmf}} \;\;,
\end{equation}
where ``cmf'' stands for comoving frame.
A computational convenience may be taken by virtue of Eq.
\eqref{eq13}.  Following Kasen, we track both IMC and DDMC particles in
velocity space.  Thus, the grid itself is logically unchanging and velocity
acts as a Lagrangian coordinate.  IMC particles move over ``velocity 
distances'' denoted with a lower case $u$.  Corresponding to physical distance
in standard IMC~\citep{fleck1971}, there is a velocity to a cell boundary $u_{b}$, 
a velocity to collision $u_{\text{col}}$, and a velocity distance to census at
the end of a time step $u_{\text{cen}}$.  Through Eq.~\eqref{eq13}, the 
velocity to a boundary can be calculated with the same formula as the physical 
distance and is consequently dependent on grid geometry.  For the $u_{\text{col}}$
and $u_{\text{cen}}$, the formulas are
\begin{equation}
\label{eq17}
u_{\text{col}}=\frac{-\ln(\xi)}{(t_{n}+t_{\min})(1-\hat{\Omega}_{p}
\cdot\vec{U_{p}}/c)(\sigma_{s,n}+\sigma_{a,n})}
\end{equation}
and
\begin{equation}
\label{eq18}
u_{\text{cen}}=c\left(\frac{t_{n}+\Delta t_{n}-t_{p}}{t_{n}+t_{\min}}\right)
\end{equation}
where $t_{p}$, $\vec{U}_{p}$, and $\hat{\Omega}_{p}$ are a particle's 
time, ``velocity position'', and direction in the lab frame, respectively.
The value $\xi\in(0,1]$ is a uniformly sampled random number.  The minimum
$u=\min(u_{\text{col}},u_{\text{cen}},u_{b})$ indicates which event occurs
at each iteration of a particle's history.

Effective absorption can be alternatively calculated with implicit capture
\citep[p.~332]{lewis1993}.  If implicit capture is used to reduce the variance of
an MC particle tally, the collision velocity only includes scattering
opacities~\citep{fleck1971},
\begin{equation}
\label{eq19}
u_{\text{col}}=\frac{-\ln(\xi)}{(t_{n}+t_{\min})(1-\hat{\Omega}_{p}
\cdot\vec{U_{p}}/c)(\sigma_{s,n}+(1-f_{n})\sigma_{a,n})} \;\;.
\end{equation}
The energy of a particle in the lab frame is reduced by
$E_{p}\rightarrow E_{p}e^{-f_{n}\frac{\nu_{0,p}}{\nu_{p}}
\sigma_{a,n}u(t_{n}+t_{\min})}$ where $\nu_{p}$ and $\nu_{0,p}$ are
the particle's lab and comoving frame frequency, respectively.

For a one dimensional, spherically symmetric shell geometry,
\begin{equation}
\label{eq20}
u_{b}=\begin{cases}
|(U_{j-1/2}^{2}-(1-\mu_{p}^{2})U_{p}^{2})^{1/2}+\mu_{p}U_{p}| \\
\;\;\;\;\text{if }\mu_{p}<-\sqrt{1-(U_{j-1/2}/U_{p})^{2}}\\
\\
(U_{j+1/2}^{2}-(1-\mu_{p}^{2})U_{p}^{2})^{1/2}-\mu_{p}U_{p}\\
\;\;\;\;\text{otherwise}
\end{cases}
\end{equation}
where $\mu_{p}$ is the projection of $\hat{\Omega}_{p}$ along the radial
coordinate.  The index $j\in\{1\ldots J\}$ denotes a velocity zone.
For a geometry in which the inner most cell has $U_{1/2}=0$, the second
case in Eq.~\eqref{eq20} must be applied for the innermost cell, $j=1$,
and $\mu_{p}\in[-1,1]$.  IMC particle position in velocity space must
be updated to have its physical position unchanged.  The censused position 
is simply maintained with
\begin{equation}
\label{eq21}
\vec{U}_{p,\text{new cen}}(t_{n}+\Delta t_{n}+t_{\min}) = \vec{U}_{p,\text{old cen}}
(t_{n}+t_{\min})=\vec{r}_{p,\text{cen}} \;\;,
\end{equation}
where $\vec{r}_{p}$ is the implied IMC particle position at the end of a
time step.  Equation~\eqref{eq21} can either be implemented before or
after the routine that advances the particle set.  We find that introducing
a time centering parameter, $\alpha_{2}$, to split the velocity position shift 
before and after transport is generally preferable.  At particle advance, the
algorithm is
\begin{enumerate}
\item $\vec{U}_{p,\text{* cen}}(t_{n}+\alpha_{2}\Delta t_{n}+t_{\min}) = 
  \vec{U}_{p,\text{old cen}}(t_{n}+t_{\min})$.
\item IMC: $\vec{U}_{p,\text{* cen}}
  \rightarrow\vec{U}_{p,*}$.
\item $\vec{U}_{p,\text{new cen}}(t_{n}+\Delta t_{n}+t_{\min}) = 
  \vec{U}_{p,*}(t_{n}+\alpha_{2}\Delta t_{n}+t_{\min})$.
\end{enumerate}
If we were transporting massive particles, then the above steps would ensure
that the particle does not move in space if it has zero velocity with respect
to the lab frame.

In many transport problems, the exact dependence of opacity on frequency is
not necessarily known.  If the fully continuous opacities are known, they may
not be practical to implement in analytic form.  One might then \textit{a priori} 
posit a group structure for the opacities that are relevant to the transport 
problem and formulate Eq.~\eqref{eq11} in terms of such a group structure.  
We may then set the group Rosseland and Planck opacities, 
$\sigma_{P,g,n}=\sigma_{R,g,n}=\sigma_{a,g,n}$.  But if each velocity cell has
its own frequency grouping and number of groups, we must constrain the resulting
equation to one particular fluid cell.  For a cell $j$, group index $g$, and a
total number of (transport) groups $G_{j}$,
\begin{multline}
\label{eq22}
(1+\hat{\Omega}_{0}\cdot\vec{U}/c)\frac{1}{c}\frac{DI_{0,g}}{Dt}+\hat{\Omega}_{0}
\cdot\nabla I_{0,g}\\
+\frac{1}{c}\hat{\Omega}_{0}\cdot\nabla\vec{U}\cdot\hat{\Omega}_{0}
(I_{0,g}-\nu_{g-1/2}I_{0,\nu_{g-1/2}}+\nu_{g+1/2}I_{0,\nu_{g+1/2}})\\
-\frac{1}{c}\hat{\Omega}_{0}\cdot\nabla\vec{U}\cdot(\mathbf{I}-\hat{\Omega}_{0}
\hat{\Omega}_{0})\cdot\nabla_{\hat{\Omega}_{0}}I_{0,g}+\frac{3}{c}\hat{\Omega}_{0}
\cdot\nabla\vec{U}\cdot\hat{\Omega}_{0}I_{0,g}\\
+(\sigma_{s,g,n}+\sigma_{a,g,n})I_{0,g}=\frac{f_{n}}{4\pi}\gamma_{g,n}
\sigma_{P,n}acT_{n}^{4}\\
+\frac{1}{4\pi}\gamma_{g,n}(1-f_{n})\sum_{g'=1}^{G_{j}}\int_{4\pi}
\sigma_{a,g',n}I_{0,g'}d\Omega_{0}'\\
+\sum_{g'=1}^{G_{j}}\int_{4\pi}\sigma_{s,n}(g'\rightarrow g,\hat{\Omega}_{0}'
\cdot\hat{\Omega}_{0})I_{0,g'}d\Omega_{0}' \;\;,
\end{multline}
where $g\in\{1\ldots G_{j}\}$, $U\in[U_{j-1/2},U_{j+1/2}]$ and $\gamma_{g,n}=
\int_{\nu_{g+1/2}}^{\nu_{g-1/2}}\sigma_{a,n}b_{0,\nu_{0},n}d\nu_{0}/\sigma_{P,n}$.  
Additionally, $I_{0,g}=\int_{\nu_{g+1/2}}^{\nu_{g-1/2}}I_{0,\nu_{0}}d\nu_{0}$ and
a higher group index corresponds to a lower frequency or equivalently a
higher wavelength.  $\mathbf{I}$ is the identity matrix.  We have made
sure to apply the group grid in the comoving frame.  Equation~\eqref{eq22}
makes use of Castor's division of the Doppler and aberration corrections 
from the photon momentum gradient.  The third term on the left-hand-side of 
Eq.~\eqref{eq22} indicates that Doppler shifting will cause particles to 
leak between groups.  Moreover, the leakage process will depend on the anisotropy 
of the radiation field as well as the velocity gradient.  We track frequency
or wavelength continuously.  Hence when an IMC particle crosses a cell boundary,
its group in the subsequent cell environment can be determined without ambiguity.
If groups in adjacent cells do not align, group intersections determine transitions
in phase space for IMC and DDMC particles~\citep{densmore2012}.
In-cell group transitions from streaming in velocity may be modeled directly
by computing a velocity distance to Doppler shift, $u_{\text{Dop}}$, and taking
$u=\min(u_{\text{col}},u_{\text{cen}},u_{b},u_{\text{Dop}})$.  The form of the velocity
to Doppler shift in a spherically symmetric outflow is
$u_{\text{Dop}}=|c(1-\nu_{g+1/2}/\nu_{p})-\vec{U}_{p}\cdot\hat{\Omega}_{p}|$ where
$\nu_{p}$ is particle lab frame frequency and use has been made of Eqs.~\eqref{eq14}
and~\eqref{eq15}.

\subsection{Comoving Frame DDMC}
\label{subsec:CFDDMC}

Integrating Eq.~\eqref{eq22} over comoving solid angle, using the
simplifications described in~\citet[pp.~102-111]{castor2004}, and assuming comoving isotropic
opacities,
\begin{multline}
\label{eq23}
\frac{1}{c}\frac{\partial\phi_{0,g}}{\partial t}+\frac{1}{c}
\nabla\cdot(\vec{U}\phi_{0,g})+\nabla\cdot\vec{F}_{0,g}+\\
(\mathbf{P}_{0,g}-\nu_{g-1/2}\mathbf{P}_{0,\nu_{g-1/2}}+\nu_{g+1/2}
\mathbf{P}_{0,\nu_{g+1/2}}):\nabla\vec{U}\\
+\left[\sigma_{s,g,n}+f_{n}\sigma_{a,g,n}+(1-\gamma_{g,n})(1-f_{n})\sigma_{a,g,n}\right]\phi_{0,g}\\
=f_{n}\gamma_{g,n}\sigma_{P,n}acT_{n}^{4}+\gamma_{g,n}(1-f_{n})\sum_{g'\not=g}^{G_{j}}
\sigma_{a,g',n}\phi_{0,g'}\\
+\sum_{g'=1}^{G_{j}}\sigma_{s,n}(g'\rightarrow g)\phi_{0,g'}
\end{multline}
where
\begin{equation} 
\phi_{0,g}=\int_{4\pi}I_{0,g}d\Omega_{0} \;\;,\nonumber
\end{equation}
\begin{equation}
\vec{F}_{0,g}=\int_{4\pi}\hat{\Omega}_{0}I_{0,g}d\Omega_{0} \;\;,\nonumber
\end{equation}
\begin{equation}
\mathbf{P}_{0,g}=\frac{1}{c}
\int_{4\pi}\hat{\Omega}_{0}\hat{\Omega}_{0}I_{0,g}d\Omega_{0}\;\;,\nonumber
\end{equation}
\begin{equation}
\mathbf{P}_{0,\nu_{g\pm1/2}}=\frac{1}{c}
\int_{4\pi}\hat{\Omega}_{0}\hat{\Omega}_{0}I_{0,\nu_{g\pm1/2}}d\Omega_{0}
\;\;,\nonumber
\end{equation}
and the expression $\mathbf{W}:\mathbf{V}$ is the trace of the matrix product
$\mathbf{W}\mathbf{V}^{T}$, i.e. $\mathbf{W}:\mathbf{V}=\sum_{k}\sum_{k'}W_{k,k'}V_{k,k'}$
\citep[p.~111]{castor2004}.  If the physical scattering is elastic, then the last term 
on the right-hand-side of Eq.~\eqref{eq23} cancels with $\sigma_{s,g,n}\phi_{0,g}$.

Equation~\eqref{eq23} is a starting point for describing multigroup DDMC
\citep{abdikamalov2012}.  The frequency groups for DDMC do not necessarily have 
to be the same as those for IMC.  If group lumping is employed over regimes of 
energy and velocity cells that are amenable to a diffusion approximation, then the DDMC 
group structure will be different (less resolved) than the IMC group structure.
The transport and diffusion groups however must complement each other over the
prescribed (user defined) energy grid.
Following Abdikamalov, we operator-split Eq.~\eqref{eq23} into a transport
component, a Doppler component, and an advection-expansion component
\citep{abdikamalov2012}:
\begin{multline}
\label{eq24}
\frac{1}{c}\left(\frac{\partial\phi_{0,g}}{\partial t}\right)_{\text{Transport}}+\nabla\cdot\vec{F}_{0,g}+\\
\left[\sigma_{s,g,n}+f_{n}\sigma_{a,g,n}+(1-\gamma_{g,n})(1-f_{n})\sigma_{a,g,n}\right]
\phi_{0,g} = \\ f_{n}\gamma_{g,n}\sigma_{P,n}acT_{n}^{4}+\gamma_{g,n}(1-f_{n})
\sum_{g'\not=g}^{G_{j}}\sigma_{a,g',n}\phi_{0,g'}\\
+\sum_{g'=1}^{G_{j}}\sigma_{s,n}(g'\rightarrow g)\phi_{0,g'} \;\;,
\end{multline}
\begin{multline}
\label{eq25}
\frac{1}{c}\left(\frac{\partial\phi_{0,g}}{\partial t}\right)_{\text{Doppler}}+\mathbf{P}_{0,g}:\nabla\vec{U}\\
=(\nu_{g-1/2}\mathbf{P}_{0,\nu_{g-1/2}}-\nu_{g+1/2}\mathbf{P}_{0,\nu_{g+1/2}}):\nabla\vec{U}
\;\;,
\end{multline}
and
\begin{equation}
\label{eq26}
\left(\frac{\partial\phi_{0,g}}{\partial t}\right)_{\text{Adv/Exp}}+\nabla\cdot(\vec{U}\phi_{0,g}) = 0 \;\;.
\end{equation}
Equations~\eqref{eq23} to~\eqref{eq26} were obtained by taking the zeroth moment
of Eq.~\eqref{eq22} in solid angle.  Multiplying Eq.~\eqref{eq22} by 
$\hat{\Omega}_{0}$, integrating over solid angle, and using Buchler's analysis
to drop insignificant terms~\citep{buchler1983}, the operator-split first moment 
equations are~\citep{castor2004,abdikamalov2012}
\begin{equation}
\label{eq27}
\frac{1}{c}\left(\frac{\partial\vec{F}_{0,g}}{\partial t}\right)_{\text{Transport}}
+c\nabla\cdot\mathbf{P}_{0,g}=-(\sigma_{s,g,n}+\sigma_{a,g,n})\vec{F}_{0,g}
\end{equation}
and
\begin{equation}
\label{eq28}
\left(\frac{\partial\vec{F}_{0,g}}{\partial t}\right)_{\text{Adv/Exp}}+\nabla\cdot(\vec{U}\vec{F}_{0,g})=0
\end{equation}
where now terms with the Fleck factor are no longer present due to isotropy.
Equations~\eqref{eq24} and~\eqref{eq27} are merely the P$_{1}$ equation set over
the Eulerian coordinate corresponding to multigroup IMC transport.  If the
intensity is only linearly anisotropic and the flux varies slowly with respect
to photon mean free time, then Eq.~\eqref{eq27} reduces to Fick's Law,
\begin{equation}
\label{eq29}
\vec{F}_{0,g}=\frac{-1}{3(\sigma_{s,g,n}+\sigma_{a,g,n})}\nabla\phi_{0,g} \;\;.
\end{equation}
Additionally, Eq.~\eqref{eq25} becomes
\begin{multline}
\label{eq30}
\frac{\partial\phi_{0,g}}{\partial t}+\frac{\nabla\cdot\vec{U}}{3}
\phi_{0,g}=\\\frac{\nabla\cdot\vec{U}}{3}(\nu_{g-1/2}\phi_{0,\nu_{g-1/2}}-
\nu_{g+1/2}\phi_{0,\nu_{g+1/2}}) \;\;.
\end{multline}
The relevant equations for the DDMC method are Eqs.~\eqref{eq24},~\eqref{eq26},
\eqref{eq28},~\eqref{eq29}, and~\eqref{eq30}.  From the MC perspective, Eqs.~\eqref{eq26}
and~\eqref{eq28} are both solved by advecting the particles along with the fluid.  
But we are transporting particles over the space of velocities in an outflow.
So to solve Eq.~\eqref{eq26} or~\eqref{eq28}, we merely leave the DDMC particles in
the cells where they are censused.  Since IMC does not call for explicit advection
of MC particles, explicit advection is DDMC specific.

Equation~\eqref{eq30} can be solved in several ways.  One might approximate
$\phi_{0,\nu_{g-1/2}}\approx\phi_{0,g-1}/\Delta\nu_{g-1}$ if $\nabla\cdot\vec{U}$ 
is positive and the local radiation field is cumulatively red-shifting, or 
$\phi_{0,\nu_{g-1/2}}\approx\phi_{0,g}/\Delta\nu_{g}$ if $\nabla\cdot\vec{U}$ is
negative and the local radiation field is blue-shifting 
($\Delta\nu_{g}=\nu_{g-1/2}-\nu_{g+1/2}$).  Incorporating the approximation for a
red-shifting field into Eq.~\eqref{eq30}, the resulting equation for the lowest
group index is a homogeneous ODE.  The solution to $g=1$ ODE can be used to calculate
the heterogeneity from between-group Doppler shifting in the remaining group equations.
Alternatively, one might apply Abdikamalov's approach of tracking particle frequency
continuously for both IMC and DDMC~\citep{abdikamalov2012}.  The homogeneous solution
to Eq.~\eqref{eq30} can be used to shift the energy weight and wavelengths of each
MC particle.  Between-group shifting is then obtained by regrouping particle histories
based on the new value of the particle frequency.  Regrouping particles after shifting
frequency accounts for the right-hand-side of Eq.~\eqref{eq30}.

To obtain the DDMC method, Eq.~\eqref{eq29} must be substituted into Eq.~\eqref{eq24}
and the result must be spatially discretized~\citep{densmore2008}.  For grey diffusion,
a DDMC cell may be adjacent to an IMC cell, the domain boundary, or another DDMC cell
\citep{densmore2007}.  For multigroup diffusion, a cell might use DDMC in one group and
IMC in another.  Using notation similar to~\cite{densmore2012}, a DDMC
equation for a cell $j$ in a fully optically-thick domain away from any domain boundaries
is
\begin{multline}
\label{eq31}
\frac{1}{c}\frac{\partial\phi_{0,j,g}}{\partial t}+\bigg[\sum_{j'}
\sigma_{j\rightarrow j',g}+\sigma_{s,j,g,n}+f_{j,n}\sigma_{a,j,g,n}\\
+(1-\gamma_{j,g,n})(1-f_{j,n})\sigma_{a,j,g,n}\bigg]\phi_{0,j,g}=
f_{j,n}\gamma_{j,g,n}\sigma_{P,j,n}acT_{j,n}^{4}+\\
\frac{1}{V_{j}}\sum_{j'}V_{j'}\sigma_{j'\rightarrow j,g}\phi_{0,j',g}+
\gamma_{j,g,n}(1-f_{j,n})\sum_{g'\not=g}^{G_{j}}\sigma_{a,j,g',n}\phi_{0,j,g'}\\
+\sum_{g'=1}^{G_{j}}\sigma_{s,j,n}(g'\rightarrow g)\phi_{0,j,g'} \;\;\;,
\end{multline}
where $j'$ is the index of a cell that shares a face with $j$, $\sigma_{j\rightarrow j',g}$
are determined from the finite volume discretization of the divergence of the flux
along with Fick's law (so they are grid geometry dependent), and
$V_{j}$ is the volume of cell $j$.  In Eq.~\eqref{eq31} and in what follows, we have dropped
the ``Transport'' subscript from $\partial\phi_{0,j,g}/\partial t$ for simplicity.  Having applied
Abdikamalov's operator split, the
usual DDMC interpretation may be given to Eq.~\eqref{eq31}.  Specifically, the ``leakage
opacities'' $\sigma_{j\rightarrow j'}$ determine how likely a DDMC particle will leak from
cell $j$ to cell $j'$, $f_{j,n}\sigma_{a,j,g,n}$ is the effective absorption opacity, and
$(1-\gamma_{j,g,n})(1-f_{j,n})\sigma_{a,j,g,n}$ determines how likely a DDMC particle will
scatter out of its current group~\citep{densmore2012}.  The source terms on the 
right-hand-side of Eq.~\eqref{eq31} are, respectively, effective thermal emission, particles
leaking into $j$ from adjacent cells, in-scattering from groups of cell $j$ into $g$,
and physical scattering.

A DDMC particle's event may be sampled from a histogram of the opacities multiplying
$\phi_{0,j,g}$ in the second term of Eq.~\eqref{eq31}~\citep{densmore2007}.  Since the
diffusion equation is kept continuous in time, each DDMC particle is thought to stream
in time~\citep{densmore2007}.  If scattering is elastic, the time to a next event is
\citep{densmore2007,densmore2012}
\begin{equation}
\label{eq31a}
\delta t_{p} = -\frac{1}{c}\frac{\ln(\xi)}{\sigma_{\text{DDMC total}}}
\end{equation}
where $\xi\in(0,1]$ is again a uniformly sampled random variable and 
$\sigma_{\text{DDMC total}}=\sum_{j'}\sigma_{j\rightarrow j',g}+f_{j,n}\sigma_{a,j,g,n}
+(1-\gamma_{j,g,n})(1-f_{j,n})\sigma_{a,j,g,n}$.

Having discretized the diffusion equation in space and frequency, we must make note of
some important ambiguities that affect the implementation of this hybrid diffusion
transport method.  The first ambiguity is DDMC particle position which we touched upon
in the introduction.  An additional ambiguity implied by Eq.~\eqref{eq31} is DDMC 
particle frequency or wavelength.  Specifically, the scattering opacity only accounts
for particles leaving their current group.  Densmore's form of multifrequency IMC-DDMC
employed a grey DDMC method for particles transporting below a threshold frequency
\citep{densmore2012}.  Applying DDMC over arbitrary group distributions is then a
generalization to opacities that may be strongly non-monotonic in frequency.  But this
slight modification to the method does not change the notion that DDMC is essentially
grey during in-group propagation.  So DDMC particles may have continuous frequency as
a property as long as it is re-sampled within the current group before being explicitly
used in the transport-diffusion algorithm~\citep{densmore2012}.  To not re-sample DDMC
frequencies at every reference in the code is to neglect the possibility that multiple
scattering processes occurred within the group.

\subsection{Moving Boundary Layer Analysis}
\label{subsec:MBLA}

If a particular DDMC cell-group, $(j,g)$, is adjacent to an IMC cell $(j',g')$ where
the groups of $g$ and $g'$ are contiguous, then one may use either a Marshak boundary
condition or the~\cite{habetler1975} asymptotic diffusion limit boundary condition.
In either case, the boundary condition for the split transport operator may be expressed as
\citep{densmore2008}
\begin{multline}
\label{eq32}
\phi_{0,g}(\vec{r}_{b},t)+\left(\frac{\lambda}
{\sigma_{a,g,n}+\sigma_{s,g,n}}\right)\vec{n}\cdot\nabla
\phi_{0,g}(\vec{r}_{b},t)=\\
2\int_{\hat{\Omega}_{0}\cdot\vec{n}<0}\int_{\nu_{g+1/2}}^{\nu_{g-1/2}}
W(|\hat{\Omega}_{0}\cdot\vec{n}|)I_{0,\nu_{0}}
(\vec{r}_{b},\hat{\Omega}_{0},t)d\nu_{0}d\Omega_{0}
\end{multline}
in the comoving frame over Eulerian coordinates, where
$\lambda\approx 0.7104$, $\vec{n}$ is the unit outward normal vector of a cell surface,
and $\vec{r}_{b}$ is a point on the cell surface~\citep{densmore2008}.  The function 
$W(\mu)=2\mu$ for an isotropic Marshak boundary condition or $W(\mu)\approx\mu+
3\mu^{2}/2$ for an approximation to Habetler's asymptotic result
\citep{densmore2007,habetler1975}.
We have tacitly assumed that diffusive scattering between groups during a boundary
crossing does not occur.  The near-equilibrium condition at the cell boundaries is
reasonable~\citep{habetler1975} when the mean free time is small compared to time
step size.  A small relative mean free time can be enforced by the same heuristic
used to determine if a cell is DDMC compatible. 


Numerical experiments in Section~\ref{sec:Heav} indicate that Eq.~\eqref{eq32}
might be insufficient for some mean free path thresholds dictating whether a
cell-group is in IMC or DDMC.  Specifically, at least in spherical geometry we have
found that applying DDMC in only cell-groups with large ($\gtrsim 10$) numbers of radial
mean free paths in a problem with a monotonic density gradient and a strong outflow
($\sim 10^{9}$cm/s) may cause an artificial depression in the hybrid radiation energy
density profiles where the code is applying Eq.~\eqref{eq32}.  It has been noted
by~\cite{densmore2008} that Eq.~\eqref{eq32} does not incorporate effects of curvature.
\cite{malvagi1991} derive an asymptotic diffusion limit boundary condition that expands
on the work of~\cite{habetler1975} by incorporating spatial curvature, spatial
variation, and opacity variation at the boundary.  For the set of Heaviside source
outflow problems along with the range of cell-group coupling heuristics we test,
the effect of spatial curvature at IMC-DDMC boundaries is found to be negligible.
Nevertheless, the work of~\cite{malvagi1991} and~\cite{case1960} may be extended
to incorporate fluid effects.  Instead of an additional curvature term in
Eq.~\eqref{eq32}, we apply a boundary layer asymptotic analysis to the O($U/c$)
comoving transport equation with frequency independence to obtain an expansion term
that may then be given a MC interpretation.  The assumption that scattering does not
occur simultaneously with DDMC-IMC boundary interactions then allows for trivial
extension to multigroup.

Neglecting O($U^{2}/c^{2}$) terms and assuming frequency independent opacity,
integrating the comoving transport equation gives
\begin{multline}
\label{eqA}
\frac{1}{c}\frac{\partial I_{0}}{\partial t}+
\hat{\Omega}_{0}\cdot\nabla I_{0}+\sigma_{t,0}I_{0}+
\frac{\vec{U}}{c}\cdot\nabla I_{0}-\\
\frac{1}{c}\hat{\Omega}_{0}
\cdot\nabla\vec{U}\cdot(\mathbf{I}-\hat{\Omega}_{0}
\hat{\Omega}_{0})\cdot\nabla_{\hat{\Omega}_{0}}I_{0}+
\frac{4}{c}\hat{\Omega}_{0}\cdot\nabla\vec{U}\cdot
\hat{\Omega}_{0}I_{0} = j_{0} \;\;,
\end{multline}
where $I_{0}=\int_{0}^{\infty}I_{0,\nu_{0}}d\nu_{0}$, $j_{0}$ is the total
frequency integrated source due to scattering events and emission and
$\sigma_{t,0}=\sigma_{s,0}+\sigma_{a,0}$ is a total opacity. Supposing
there exists some spatial surface denoted by $b$, we now make use of the
homologous outflow Eq.~\eqref{eq13} to obtain
\begin{multline}
\label{eqB}
\frac{1}{c}\frac{\partial I_{0}}{\partial t}+
\mu\frac{\partial I_{0}}{\partial z}+\mathcal{L}I_{0}+
(\hat{\Omega}_{0}\cdot\nabla)_{\perp}I_{0}+\\
\sigma_{t,0}I_{0}+\frac{\vec{r}}{ct_{f}}
\cdot\nabla I_{0}+\frac{4}{ct_{f}}I_{0} = j_{0} \;\;,
\end{multline}
where $\mu$ is the projection of the comoving angle onto an axis
$z$ aligned orthogonal to a plane tangent to surface $b$, the linear
operator $\mathcal{L}$ accounts for the change in the directional 
derivative, $\hat{\Omega}_{0}\cdot\nabla$, due to non-trivial coordinate 
curvature, $(\hat{\Omega}_{0}\cdot\nabla)_{\perp}$ is the projection
of the directional derivative orthogonal to $z$~\citep{malvagi1991},
and the ``fluid time'' $t_{f}=t_{n}+t_{\min}$ upon implementation.
Now we take the usual step of postulating a parameter $\varepsilon
\ll 1$ such that~\citep{habetler1975,malvagi1991}
\begin{subequations}
\label{eqC}
\begin{align}
&c\rightarrow c/\varepsilon\;\;,\\
&\sigma_{t,0}\rightarrow\sigma_{t,0}/\varepsilon\;\;,\\
&\sigma_{a,0}\rightarrow\varepsilon\sigma_{a,0}\;\;,\\
&t_{f}\rightarrow t_{f}/\varepsilon\;\;,\label{eqCd}\\
&\vec{r}_{b}\rightarrow\vec{r}_{b}/\varepsilon\;\;,\label{eqCe}\\
&s = z/\varepsilon\;\;,\\
&\vec{U}(\vec{r},t_{f})\approx\vec{U}(\vec{r}_{b},t_{f})\;\;,
\end{align}
\end{subequations}
where $\vec{r}_{b}$ is a location on surface $b$ and the rescaling
in Eq.~\eqref{eqC} permits treating the parameters as O(1).  We
have scaled the surface coordinate and characteristic fluid time
to be large in Eqs.~\eqref{eqCd} and~\eqref{eqCe}.  If $\hat{r}_{b}$
and $\hat{z}$ are unit vectors of the surface coordinate and
$z$, respectively, then incorporating Eq.~\eqref{eqC} into
Eq.~\eqref{eqB} gives
\begin{multline}
\label{eqD}
\frac{\varepsilon^{2}}{c}\frac{\partial I_{0}}{\partial t}+
\left[\mu+\varepsilon(\hat{r}_{b}\cdot\hat{z})\frac{r_{b}}{ct_{f}}\right]
\frac{\partial I_{0}}{\partial s}+\varepsilon\left[\mathcal{L}+
\varepsilon\frac{r_{b}}{ct_{f}}\mathcal{L}_{\hat{r}}\right]I_{0}+\\
\varepsilon\left[(\hat{\Omega}_{0}\cdot\nabla)_{\perp}+
\varepsilon\frac{r_{b}}{ct_{f}}(\hat{r}_{b}\cdot\nabla)_{\perp}\right]I_{0}+
\sigma_{t,0}I_{0}+\frac{4\varepsilon^{2}}{ct_{f}}I_{0} = \varepsilon j_{0}
\end{multline}
and
\begin{equation}
\label{eqE}
j_{0}=\left(\frac{\sigma_{t,0}}{\varepsilon}+\varepsilon\sigma_{a,0}\right)
\frac{1}{4\pi}\int_{4\pi}I_{0}d\Omega_{0}+\varepsilon q \;\;,
\end{equation}
where $q$ is external or thermal sources, $r_{b}=|\vec{r}_{b}|$, and
the opacities have been assumed isotropic.  We have defined
$\hat{r}_{b}\cdot\nabla=(\hat{r}_{b}\cdot\hat{z})\partial/\partial z
+(\hat{r}_{b}\cdot\nabla)_{\perp}+\mathcal{L}_{\hat{r}}$ as a means
of tracking the change with respect to the ballistic fluid
trajectories through curved coordinates in a fashion
analogous to the streaming term for photon trajectories.  At least for
spherical symmetry, Eq.~\eqref{eqCe} implies $\mathcal{L}\rightarrow
\varepsilon\mathcal{L}$ in agreement with intuition.  Incorporating 
Eq.~\eqref{eqE} into Eq.~\eqref{eqD} and grouping O($\varepsilon^{2}$)
terms on the right hand side,
\begin{multline}
\label{eqF}
\left[\mu+\varepsilon(\hat{r}_{b}\cdot\hat{z})\frac{r_{b}}{ct_{f}}\right]
\frac{\partial I_{0}}{\partial s}+\varepsilon\mathcal{L}I_{0}+\\
\varepsilon(\hat{\Omega}_{0}\cdot\nabla)_{\perp}I_{0}+
\sigma_{t,0}I_{0}=\frac{\sigma_{t,0}}{4\pi}\int_{4\pi}I_{0}d\Omega_{0}'
+O(\varepsilon^{2}) \;\;.
\end{multline}
It is interesting to note that Eq.~\eqref{eqC} prescribes scalings
that make changes in $I_{0}$ from curvature and surface variation
along the ballistic fluid parcel trajectories at surface $b$ an
O($\varepsilon^{2}$) effect.  This is useful, as we only consider
O($\varepsilon$) effects in the construction of the boundary condition.
We now neglect surface variations and curvature in the following analysis
as these conditions have been thoroughly examined by~\cite{malvagi1991}.
Following prior authors, we also separate $I_{0}$ into a boundary layer
solution, $I_{0,b}$, and an interior solution, $I_{0,i}$ such that $I_{0}=
I_{0,b}+I_{0,i}$ and $\lim_{s\rightarrow\infty}I_{0,b}=0$~\citep{habetler1975,
malvagi1991}.  For $I_{0,b}$, Eq.~\eqref{eqF} then reduces further to
\begin{multline}
\label{eqG}
\left[\mu+\varepsilon(\hat{r}_{b}\cdot\hat{z})\frac{r_{b}}{ct_{f}}\right]
\frac{\partial I_{0,b}}{\partial s}+\sigma_{t,0}I_{0,b}=\\
\frac{\sigma_{t,0}}{4\pi}\int_{4\pi}I_{0,b}d\Omega_{0}'
+O(\varepsilon^{2}) \;\;.
\end{multline}
The boundary and interior solutions may be expanded in the small
parameter $\varepsilon$ as $I_{0,(b,i)}=\sum_{m=0}^{\infty}I_{0,(b,i)}^{(m)}
\varepsilon^{m}$.  Incorporating the $\varepsilon$-expansion into
Eq.~\eqref{eqG}, balancing $\varepsilon^{0}$ and $\varepsilon^{1}$
coefficients, and integrating over the azimuthal angle about $z$,
the O(1) and O($\varepsilon$) equations are
\begin{equation}
\label{eqH}
\mu\frac{\partial\tilde{I}_{0,b}^{(0)}}{\partial s}+\sigma_{t,0}
\left(\tilde{I}_{0,b}^{(0)}-\frac{1}{2}\int_{-1}^{1}\tilde{I}_{0,b}^{(0)}
(\mu')d\mu'\right)=0
\end{equation}
\begin{multline}
\label{eqI}
\mu\frac{\partial\tilde{I}_{0,b}^{(1)}}{\partial s}+\sigma_{t,0}
\left(\tilde{I}_{0,b}^{(1)}-\frac{1}{2}\int_{-1}^{1}\tilde{I}_{0,b}^{(1)}
(\mu')d\mu'\right)=\\-(\hat{r}_{b}\cdot\hat{z})
\frac{r_{b}}{ct_{f}}\frac{\partial\tilde{I}_{0,b}^{(0)}}
{\partial s}
\end{multline}
where $\tilde{I}_{0,(b,i)}^{(m)}=\int_{0}^{2\pi}I_{0,(b,i)}^{(m)}d\omega$
and $\omega$ is the azimuthal angle.
Instead of the curvature and spatial variation terms, the
heterogeneity of the O($\epsilon$) equation is an expansion
term.  Supposing the boundary intensity at $s=0$ is known,
matching the asymptotic orders gives~\citep{malvagi1991}
\begin{equation}
\label{eqJ}
\tilde{I}_{0,b}^{(0)}(s=0,\mu,t)=F(\vec{r}_{b},\mu,t)-\frac{1}{2}
\phi_{0,i}^{(0)}(\vec{r}_{b},t)
\end{equation}
and
\begin{equation}
\label{eqK}
\tilde{I}_{0,b}^{(1)}(s=0,\mu,t)=-\frac{1}{2}
\left(\phi_{0,i}^{(1)}(\vec{r}_{b},t)-\frac{\mu}{\sigma_{0,t}}
\left.\frac{\partial\phi_{0,i}^{(0)}}{\partial z}\right|_{\vec{r}_{b}}
\right)
\end{equation}
as boundary conditions, where $\mu>0$ is the magnitude of
the angular projection into the diffusive domain along axis $z$,
$\phi_{0,(b,i)}^{(m)}=\int_{4\pi}I_{0,(b,i)}^{(m)}d\Omega_{0}$
and $F(\vec{r}_{b},\mu,t)=\int_{0}^{2\pi}I_{0}(\vec{r}_{b},
\hat{\Omega}_{0},t)d\omega$.  Eq.~\eqref{eqH} along with Eq.
\eqref{eqJ} is a form of the standard half-space albedo
problem examined by~\cite{case1960} and~\cite{larsen1973}.
The solution to Eq.~\eqref{eqH} is~\citep{malvagi1991}
\begin{equation}
\label{eqL}
\tilde{I}_{0,b}^{(0)}=k_{+}^{(0)}+\int_{0}^{1}k^{(0)}(\varpi)
\varphi_{\varpi}(\mu)e^{-\sigma_{t,0}s/\varpi}d\varpi \;\;,
\end{equation}
where $k_{+}^{(0)}$ is a constant, $\varpi$ is an eigenvalue
of the singular eigenfunction $\varphi_{\varpi}(\mu)$ (the
eigenfunction is formally a distribution),
and $k^{(0)}(\varpi)$ is a function determined by the
orthogonalities, Eqs.~\eqref{eqO} and~\eqref{eqP} below
\citep{habetler1975}.  The constant,
$k_{+}^{(0)}$, constitutes the ``discrete'' component of the
solution~\citep{habetler1975} and is the limiting behavior
of a more general discrete eigenvalue solution having taken
$\sigma_{0,a}\ll\sigma_{0,s}$ with $\varepsilon$.
The distribution, $\varphi_{\varpi}(\mu)$, is given by
\begin{equation}
\label{eqM}
\varphi_{\varpi}(\mu)=\frac{\varpi}{2}P:\frac{1}{\varpi-\mu}
+\lambda(\varpi)\delta(\varpi-\mu) \;\;,
\end{equation}
where $\delta(\varpi-\mu)$ is the Dirac delta function and
$\lambda(\varpi)=1-\varpi\tanh^{-1}(\varpi)$ ensures
a normalization of $\int_{-1}^{1}\varphi_{\varpi}(\mu)d\mu=1$
for all $\varpi\in[0,1]$.  The $P:$ is merely a notational
device to indicate the principal value is taken upon integration
\citep{habetler1975,case1960}.  \cite{case1960} rigorously
proves that a function, $H(\mu)$, may be found such that
\begin{equation}
\label{eqN}
\int_{0}^{1}\mu H(\mu)\varphi_{\varpi}(\mu)d\mu = 0
\end{equation}
when $\tilde{I}_{0,b}^{(0)}$ satisfies a
H\"{o}lder condition.  It turns out $H(\mu)$ is Chandrasekhar's
H-function~\citep{malvagi1991}. Making use of the
Poincar\'{e}-Bertrand formula~\citep{hang2009} and Eq.~\eqref{eqN},
the orthogonalities~\citep{malvagi1991}
\begin{equation}
\label{eqO}
\int_{-1}^{1}\mu\varphi_{\varpi'}(\mu)\varphi_{\varpi}(\mu)d\mu=
N(\varpi)\delta(\varpi-\varpi')
\end{equation}
and
\begin{equation}
\label{eqP}
\int_{0}^{1}\mu H(\mu)\varphi_{\varpi'}(\mu)\varphi_{\varpi}(\mu)
d\mu= N(\varpi)H(\varpi)\delta(\varpi-\varpi') \;\;,
\end{equation}
where $N(\varpi)=\varpi(\lambda(\varpi)^{2}+(\pi\varpi/2)^{2})$,
must hold.  Incorporating Eq.~\eqref{eqM} into Eq.~\eqref{eqL} and
Eq.~\eqref{eqL} into the boundary condition Eq.~\eqref{eqJ},
Eqs.~\eqref{eqN} and~\eqref{eqP} indicate that multiplying the
result by $\mu H(\mu)$ or $\mu H(\mu)\varphi_{\varpi'}(\mu)$ and
integrating over $\mu\in[0,1]$ yields
\begin{equation}
\label{eqQ}
k_{+}^{(0)}=\frac{\sqrt{3}}{2}\int_{0}^{1}\mu H(\mu)F(\vec{r}_{b},
\mu)d\mu - \frac{1}{2}\phi_{0,i}(\vec{r}_{b})
\end{equation}
or
\begin{equation}
\label{eqR}
k^{(0)}(\varpi)=\frac{1}{N(\varpi)H(\varpi)}\int_{0}^{1}\mu H(\mu)
\varphi_{\varpi}(\mu)F(\vec{r}_{b},\mu)d\mu \;\;,
\end{equation}
respectively~\citep{malvagi1991} ($\int_{0}^{1}\mu H(\mu)=2/\sqrt{3}$
and we have dropped $t$ as an argument).  As $s\rightarrow\infty$,
Eq.~\eqref{eqI} becomes homogeneous; then as $s\rightarrow\infty$,
the O($\varepsilon$) solution $\tilde{I}_{0,b}^{(1)}$ must tend to
some constant, $k_{+}^{(1)}$, as well.  Now using the boundary
condition~\eqref{eqL}, $k_{+}^{(1)}$ may be found in a similar
manner to $k_{+}^{(0)}$ as
\begin{multline}
\label{eqS}
k_{+}^{(1)}=\frac{\sqrt{3}}{2}\int_{0}^{1}\mu H(\mu)
\left[-\frac{1}{2}\left(\phi_{0,i}^{(1)}(\vec{r}_{b})-
\frac{\mu}{\sigma_{0,t}}\left.\frac{\partial\phi_{0,i}^{(0)}}{\partial z}
\right|_{\vec{r}_{b}}\right)\right]d\mu\\
-\int_{0}^{\infty}\int_{-1}^{1}\beta(s,\mu)(\hat{r}_{b}\cdot\hat{z})
\frac{r_{b}}{ct_{f}}\frac{\partial\tilde{I}_{0,b}^{(0)}}{\partial s}d\mu
\,ds
\end{multline}
where a considerable variational analysis by~\cite{malvagi1991} gives
\begin{equation}
\label{eqT}
\beta(s,\mu)=1+\frac{3}{2}(\mu+\sigma_{0,t}s)-\left(1+\frac{3}{2}\mu
\right)e^{-\sigma_{0,t}s/\mu}\Theta(-\mu)\;\;,
\end{equation}
and $\Theta$ is the Heaviside function.  Roughly speaking, to
obtain Eq.~\eqref{eqT},~\cite{malvagi1991} construct a linear
functional for $k_{+}^{(1)}$ with Lagrange multipliers as an estimate
to $k_{+}^{(1)}$, find an adjoint transport solution~\cite[p.~47]
{lewis1993} and $\mu$ times the adjoint solution as the appropriate 
multipliers, and incorporate constants as trial functions for
$\tilde{I}_{b}^{(1)}$ and the adjoint solution.

Now we may use the boundary layer constraint,
$\lim_{s\rightarrow\infty}\tilde{I}_{0,b}=0$; $k_{+}^{0}$ and
$k_{+}^{(1)}$ must vanish.  Equations~\eqref{eqS}
becomes~\citep{malvagi1991}
\begin{multline}
\label{eqU}
\phi_{0,i}^{(1)}(\vec{r}_{b})+\frac{\lambda}{\sigma_{t}}\hat{z}\cdot\nabla
\phi_{0,i}^{(0)}=2(\hat{r}_{b}\cdot\hat{z})\frac{r_{b}}{ct_{f}}
\int_{0}^{\infty}\ldots\\\left(\int_{-1}^{1}\beta(\eta,\mu)\int_{0}^{1}
\frac{1}{\varpi}k(\varpi)\varphi_{\varpi}(\mu)e^{-\eta/\varpi}
d\varpi\,d\mu\right)\,d\eta \;\;,
\end{multline}
where $\lambda=\sqrt{3}\int_{0}^{1}\mu^{2}H(\mu)d\mu/2$ and
$\sigma_{t,0}s=\eta$.  We have incorporated the
eigenvalue form of $\tilde{I}_{0,b}^{(0)}$ into Eq.~\eqref{eqU}.
Multiplying Eq.~\eqref{eqU} by $\varepsilon$, adding the
result to $\phi_{0,i}^{(0)}=\sqrt{3}\int_{0}^{1}\mu H(\mu)
F(\vec{r}_{b},\mu)d\mu$, reintroducing the interior
solution $\phi_{0,i}=\phi_{0,i}^{(0)}+\varepsilon\phi_{0,i}^{(1)}+
O(\varepsilon^{2})$ , and reverting the $\varepsilon$-scalings
from Eq.~\eqref{eqC} gives
\begin{multline}
\label{eqV}
\phi_{0,i}(\vec{r}_{b},t)+\frac{\lambda}{\sigma_{t,0}}\hat{z}\cdot\nabla
\phi_{0,i}=2(\hat{r}_{b}\cdot\hat{z})\frac{r_{b}}{ct_{f}}
\int_{0}^{\infty}\ldots\\\left(\int_{-1}^{1}\beta(\eta,\mu)\int_{0}^{1}
\frac{1}{\varpi}k(\varpi)\varphi_{\varpi}(\mu)e^{-\eta/\varpi}
d\varpi\,d\mu\right)\,d\eta\\
+\sqrt{3}\int_{0}^{1}\mu H(\mu)F(\vec{r}_{b},\mu,t)d\mu \;\;,
\end{multline}
which should be correct to O($\varepsilon^{2}$).
We find
\begin{multline}
\label{eqW}
\int_{0}^{\infty}\int_{-1}^{1}\beta(\eta,\mu)\varphi_{\varpi}(\mu)
e^{-\eta/\varpi}d\mu d\eta=\\
\varpi\left(1+\frac{3}{2}\varpi\right)
-\frac{\varpi^{2}}{2}(1+3\varpi)\ln\left(\frac{1+\varpi}{\varpi}\right)
\\
+\frac{\varpi^{2}}{2(1+\varpi)}\left(\frac{5}{2}+3\varpi\right)\\
\equiv\varpi h(\varpi) \;\;,
\end{multline}
so
\begin{multline}
\label{eqX}
\int_{0}^{1}\frac{k^{(0)}(\varpi)}{\varpi}
\int_{0}^{\infty}\int_{-1}^{1}\beta(\eta,\mu)\varphi_{\varpi}(\mu)
e^{-\eta/\varpi}\,d\mu\,d\eta\,d\varpi=\\
\int_{0}^{1}h(\varpi)k^{(0)}(\varpi)d\varpi \;\;.
\end{multline}
Incorporating Eq.~\eqref{eqR}, using $\sqrt{3}\mu H(\mu)\approx 2W(\mu)$
which is a corollary of the variational derivation of $\beta(s,\mu)$
by~\cite{malvagi1991}, and defining
\begin{equation}
\label{eqY}
G_{U}(\mu)=1+\frac{2}{c}\hat{z}\cdot\vec{U}(\vec{r}_{b},t_{f})
\int_{0}^{1}\frac{h(\varpi)\varphi_{\varpi}(\mu)}{(2+3\varpi)N(\varpi)}
d\varpi\;\;,
\end{equation}
Eq.~\eqref{eqV} becomes
\begin{equation}
\label{eqZ}
\phi_{0,i}(\vec{r}_{b},t)+\frac{\lambda}{\sigma_{t,0}}\hat{z}\cdot\nabla
\phi_{0,i}=
2\int_{0}^{1}W(\mu)G_{U}(\mu)F(\vec{r}_{b},\mu,t)d\mu \;\;.
\end{equation}
Notwithstanding the factor $G_{U}$, the form of Eq.~\eqref{eqZ} is
fortunately similar to Eq.~\eqref{eq32}.  To evaluate the integral
in Eq.~\eqref{eqZ}, we incorporate Eq.~\eqref{eqM} for $\varphi_{\varpi}(\mu)$
to obtain
\begin{multline}
\label{eqAA}
\int_{0}^{1}\frac{h(\varpi)\varphi_{\varpi}(\mu)}{(2+3\varpi)N(\varpi)}d\varpi=
\lambda(\mu)\frac{h(\mu)}{(2+3\mu)N(\mu)}+\\
\frac{1}{2}\int_{0}^{1}\left(\varpi\frac{h(\varpi)}{(2+3\varpi)N(\varpi)}-
\mu\frac{h(\mu)}{(2+3\mu)N(\mu)}\right)
\frac{d\varpi}{(\varpi-\mu)}\\+
\frac{\mu}{2}\frac{h(\mu)}{(2+3\mu)N(\mu)}\ln\left(\frac{1-\mu}{\mu}\right)
\;\;,
\end{multline}
where, following~\cite{malvagi1991}, we have converted the principal
value integration to a nonsingular form amenable to quadrature.

Equation~\eqref{eqAA} along with the form of the functions
$\lambda(\mu)$, $N(\mu)$, and $h(\mu)$ reveal the leading order
behavior of the angular dependence in $G_{U}$.  For the case of
$\mu\rightarrow 0$: $\lambda(\mu)\rightarrow 1$, $N(\mu)\rightarrow\mu$
and $h(\mu)\rightarrow 1$.  Hence the first term on the right-hand-side
of Eq.~\eqref{eqAA} tends to $0.5/\mu$ as $\mu\rightarrow 0$.  From the
last term on the right-hand-side of Eq.~\eqref{eqAA}, the next order
of divergence as $\mu\rightarrow 0$ is logarithmic.  The
remaining terms tend to a bounded integral over $\varpi$ as
$\mu\rightarrow 0$.  We find that the behavior in $\mu$ of Eq.~\eqref{eqAA}
is well approximated by $C_{1}/\mu-C_{2}\mu$ where $C_{1}$ and $C_{2}$
are positive constants.  In Section~\ref{sec:Heav}, we
test $0.5c(G_{U}-1)/(\hat{z}\cdot\vec{U})=0.55/\mu-1.25\mu$ for a
three mean free path threshold between IMC and DDMC and
$0.5c(G_{U}-1)/(\hat{z}\cdot\vec{U})=0.6/\mu-1.25\mu$ for a ten
mean free path threshold between IMC and DDMC.  These choices
are seen to be good approximations to the quadrature expressed in
Eq.~\eqref{eqAA}.

Finally, in Eq.~\eqref{eqZ} we replace $\hat{z}$ with $\vec{n}$,
drop the subscript $i$, replace $\mu$ with $|\hat{\Omega}_{0}\cdot
\vec{n}|$, and extend to multigroup to obtain
\begin{multline}
\label{eqAB}
\phi_{0,g}(\vec{r}_{b},t)+\left(\frac{\lambda}
{\sigma_{a,g,n}+\sigma_{s,g,n}}\right)\vec{n}\cdot\nabla
\phi_{0,g}(\vec{r}_{b},t)=\\
2\int_{\hat{\Omega}_{0}\cdot\vec{n}<0}\int_{\nu_{g+1/2}}^{\nu_{g-1/2}}
W(|\hat{\Omega}_{0}\cdot\vec{n}|)G_{U}(|\hat{\Omega}_{0}\cdot\vec{n}|)
\ldots\\
I_{0,\nu_{0}}(\vec{r}_{b},\hat{\Omega}_{0},t)d\nu_{0}d\Omega_{0}\;\;.
\end{multline}
when $\vec{U}=0$, the standard diffusion limit boundary condition
Eq.~\eqref{eq32} is obtained.  As is seen in Section~\ref{sec:Heav},
some form of the new factor, $G_{U}$, must be implemented at IMC-DDMC
boundaries to furnish satisfactory agreement between the hybrid method
and pure IMC over a grid moving at relativistic speed.


\subsection{Mixed Frame IMC-DDMC}
\label{subsec:MFIMCDDMC}

We now turn to the discussion of hybridizing IMC and DDMC in velocity space (cells)
and frequency space (groups).
To incorporate Eq.~\eqref{eqAB} into the DDMC routine, the second term on the
left-hand-side can be finite differenced and incorporated into the discretized
form of Eq.~\eqref{eq29}.  The resulting expression for $\vec{F}_{0,g}$ may then
be incorporated into the discrete form of Eq.~\eqref{eq24} to yield
\begin{multline}
\label{eq33}
\frac{1}{c}\frac{\partial\phi_{0,j,g}}{\partial t}+\bigg[\sum_{j''\not=j'}
\sigma_{j\rightarrow j'',g}+\sigma_{b(j,j'),g}+\sigma_{s,j,g,n}+f_{j,n}\sigma_{a,j,g,n}\\
+(1-\gamma_{j,g,n})(1-f_{j,n})\sigma_{a,j,g,n}\bigg]\phi_{0,j,g}=
f_{j,n}\gamma_{j,g,n}\sigma_{P,j,n}acT_{j,n}^{4}+\\
\frac{1}{V_{j}}\sum_{j''\not=j}V_{j''}\sigma_{j''\rightarrow j,g}\phi_{0,j'',g}+
\gamma_{j,g,n}(1-f_{j,n})\sum_{g'\not=g}^{G_{j}}\sigma_{a,j,g',n}\phi_{0,j,g'}\\
+\frac{1}{V_{j}}\int_{A_{b(j,j')}}\int_{\hat{\Omega}_{0}\cdot\vec{n}<0}
\int_{\nu_{g+1/2}}^{\nu_{g-1/2}}\ldots\\ P_{b(j,j')}(|\hat{\Omega}_{0}\cdot\vec{n}|)
|\hat{\Omega}_{0}\cdot\vec{n}|I_{0,g}d\nu_{0}d\Omega_{0}d^{2}\vec{r}\\
+\sum_{g'=1}^{G_{j}}\sigma_{s,j,n}(g'\rightarrow g)\phi_{0,j,g'} \;\;\;,
\end{multline}
where the $b(j,j')$ subscript indicates the boundary between DDMC cell $j$
and IMC cell $j'$, $\sigma_{b(j,j'),g}$ is the modified leakage opacity resulting 
from Eq.~\eqref{eqAB}, $j''$ denotes cells adjacent to $j$ with a diffusive group
$g$, $A_{b(j,j')}$ is the area of the boundary between cells $j$ and $j'$, and
$P_{b(j,j')}$ may be interpreted as a transmission probability for an IMC
particle in cell $j'$ incident on cell $j$~\citep{densmore2007}.  We have only
converted one cell-group adjacent to $(j,g)$ to IMC in Eq.~\eqref{eq33} but
in principle any of the leakage opacities and sources could be replaced by
$\sigma_{b(j,j'')}$ and IMC boundary transmission sources.  In spherically
symmetric geometry over a velocity grid, the forms of $\sigma_{j\rightarrow j'',g}$,
$\sigma_{b(j,j'),g}$, and $P_{b(j,j')}$ are~\citep{abdikamalov2012}
\begin{multline}
\label{eq34}
\sigma_{j\rightarrow j'',g}=\\
\begin{cases}
\displaystyle\frac{2U_{j-1/2}^{2}}{(t_{n}+t_{\min})^{2}\Delta(U^{3})_{j}}
\left(\frac{1}{\sigma_{j-1/2,g,n}^{-}\Delta U_{j-1}+
\sigma_{j-1/2,g,n}^{+}\Delta U_{j}}\right) \\\;\;,\;\;j''=j-1\\
\\
\displaystyle\frac{2U_{j+1/2}^{2}}{(t_{n}+t_{\min})^{2}\Delta(U^{3})_{j}}
\left(\frac{1}{\sigma_{j+1/2,g,n}^{-}\Delta U_{j}+
\sigma_{j+1/2,g,n}^{+}\Delta U_{j+1}}\right) \\\;\;,\;\;j''=j+1
\end{cases}
\\\;\;\;,
\end{multline}
\begin{multline}
\label{eq35}
\sigma_{b(j,j'),g}=\\
\begin{cases}
\displaystyle\frac{2U_{j-1/2}^{2}}{(t_{n}+t_{\min})\Delta(U^{3})_{j}}
\left(\frac{1}{\sigma_{j-1/2,g,n}^{+}(t_{n}+t_{\min})\Delta U_{j}+
2\lambda}\right)\\\;\;,\;\;j'=j-1\\
\\
\displaystyle\frac{2U_{j+1/2}^{2}}{(t_{n}+t_{\min})\Delta(U^{3})_{j}}
\left(\frac{1}{\sigma_{j+1/2,g,n}^{-}(t_{n}+t_{\min})\Delta U_{j}+
2\lambda}\right) \\\;\;,\;\;j'=j+1
\end{cases}
\\\;\;\;,
\end{multline}
and
\begin{equation}
\label{eq36}
P_{b(j,j'),g}(\mu)=\\
\begin{cases}
\displaystyle
\frac{4(1+3\mu/2)G_{U_{j-1/2}}(\mu)}
{3\sigma_{j-1/2,g,n}^{+}(t_{n}+t_{\min})\Delta U_{j}+6\lambda}\\
\;\;,\;\;j'=j-1\\
\\
\displaystyle
\frac{4(1+3\mu/2)G_{U_{j+1/2}}(\mu)}
{3\sigma_{j+1/2,g,n}^{-}(t_{n}+t_{\min})\Delta U_{j}+6\lambda}\\
\;\;,\;\;j'=j+1
\end{cases}
\end{equation}
respectively, where $\Delta U_{j}=U_{j+1/2}-U_{j-1/2}$ is the radial velocity
width, $\Delta(U^{3})_{j}=U_{j+1/2}^{3}-U_{j-1/2}^{3}$.  Following~\cite{densmore2007},
the $\sigma_{j\pm1/2,g,n}=\sigma_{j\pm1/2,a,g,n}+\sigma_{j\pm1/2,s,g,n}$ are evaluated
with cell-edge temperature while the $+$ ($-$) superscript indicates the remaining
material properties are evaluated on the outer (inner) side of the cell edge with
respect to index $j$.  The asymptotic diffusion-limit $W(\mu)$
has been used to determine $P_{b(j,j'),g}(\mu)$.  It can be shown that the
expressions in Eqs.~\eqref{eq34}-\eqref{eq36} tend to the appropriate planar
geometry forms when the inner radius of cell $j$ is large with respect to the width
of cell $j$.  

Since DDMC tracks comoving radiation energy, the interface conditions are in
the IMC-DDMC comoving frame as well.  Spatial hybridization of IMC and DDMC
in the comoving frame is evident from Eq.~\eqref{eq33}.  With a hybridized
IMC-DDMC method, the DDMC particles are advected along with the material
while the IMC particles are not.  Since an IMC particle may be moved over the
velocity grid past cell bounds, a DDMC zone may eventually advect into it.
In our outflow tests, we find that the best treatment of this occurrence is to
stop the IMC particle at the DDMC cell boundary and allow the diffuse albedo
condition described by Eq.~\eqref{eq36} to determine the particle's admission into
the DDMC region.  So the split velocity position shift algorithm for IMC particle
$p$ is now:
\begin{enumerate}
\item Find current fluid cell $j$ from $\vec{U}_{p,\text{old cen}}$ and expected
  fluid cell $j'$ from $\vec{U}_{p,\text{* cen}}
  (t_{n}+\alpha_{2}\Delta t_{n}+t_{\min}) = \vec{U}_{p,\text{old cen}}(t_{n}+t_{\min})$.
\begin{enumerate}
\item For the first diffusive cell $j''$ that passes particle $p$ in the
  span of $\alpha_{2}\Delta t$, set particle $p$'s velocity position to the point
  it would pass on the surface of $j''$, $\vec{U}_{p,\text{* cen}}=\vec{U}_{p,b}$.
\item If there are no DDMC cells that pass $p$ in $\alpha_{2}\Delta t$, then
  set $j=j'$ and use formula from step (1) for $\vec{U}_{p,\text{* cen}}$.
\end{enumerate}
\item IMC-DDMC: $\vec{U}_{p,\text{* cen}}\rightarrow\vec{U}_{p,*}$
\item If $p$ has become a DDMC particle, the position update is finished.
\item Otherwise, use $\vec{U}_{p,\text{new cen}}(t_{n}+\Delta t_{n}+t_{\min}) = 
  \vec{U}_{p,*}(t_{n}+\alpha_{2}\Delta t_{n}+t_{\min})$ to find fluid cell $j'$.
\begin{enumerate}
\item For the first diffusive cell $j''$ that passes particle $p$ in the
  span of $(1-\alpha_{2})\Delta t$, set particle $p$'s velocity position to the point
  it would pass on the surface of $j''$, $\vec{U}_{p,\text{new cen}}=\vec{U}_{p,b}$.
\item If there are no DDMC cells that pass $p$ in $(1-\alpha_{2})\Delta t$, then
  set $j=j'$ and use formula from step (4) for $\vec{U}_{p,\text{new cen}}$.
\end{enumerate}
\end{enumerate}

Having discussed hybridization of IMC with DDMC at velocity grid boundaries, it
remains to discuss the treatment of hybrid scattering and
the effect non-uniform group structuring at grid boundaries.  The hybrid scattering
follows simply from both Abdikamalov and Densmore's multifrequency IMC-DDMC methods.
We may replace one of the DDMC scattering group source terms in Eq.~\eqref{eq31} or Eq.
\eqref{eq33} with an IMC scattering source:
\begin{multline}
\label{eq37}
\gamma_{j,g,n}(1-f_{j,n})\sigma_{a,j,g',n}\phi_{0,j,g'}\rightarrow\\
\frac{\gamma_{j,g,n}(1-f_{j,n})}{V_{j}}\int_{V_{j}}\int_{4\pi}
\int_{\nu_{g'+1/2}}^{\nu_{g'-1/2}}\sigma_{a,j,n}I_{0,\nu_{0}}d\nu_{0}d\Omega_{0}d^{3}\vec{r}
\end{multline}
for effective scattering and
\begin{multline}
\label{eq38}
\sigma_{s,j,n}(g'\rightarrow g)\phi_{0,j,g'}\rightarrow\\
\frac{1}{V_{j}}\int_{V_{j}}\int_{4\pi}\int_{4\pi}\int_{\nu_{g+1/2}}^{\nu_{g-1/2}}
\int_{\nu_{g'+1/2}}^{\nu_{g'-1/2}}\frac{\nu_{0}}{\nu_{0}'}\ldots\\
\sigma_{s,j,n}(\nu_{0}'\rightarrow\nu_{0},\hat{\Omega}_{0}'\cdot\hat{\Omega}_{0})
I_{0,\nu_{0}'}d\nu_{0}'d\nu_{0}d\Omega_{0}'d\Omega_{0}d^{3}\vec{r}
\end{multline}
for physical scattering.  For the non-uniform group structuring at grid boundaries,
we extend the Planck averaging formula presented by~\cite{densmore2012} to
\begin{multline}
\label{eq39}
\tilde{\sigma}_{j\rightarrow j',g} = \\
\left(\frac{\sum_{g_{D}'}b_{j,g\leftrightarrow g_{D}',n}}{b_{j,g,n}}\right)
\sigma_{j\rightarrow j',g}
+\left(\frac{\sum_{g_{T}'}b_{j,g\leftrightarrow g_{T}',n}}{b_{j,g,n}}\right)
\sigma_{b(j,j'),g}
\end{multline}
where $g_{D}'$ is a DDMC group index in cell $j'$, $g_{T}'$ is an IMC group index in
cell $j'$, $b_{j,g,n}=\int_{\nu_{g+1/2}}^{\nu_{g-1/2}}b_{0,\nu_{0}}(T_{j,n})d\nu_{0}$, and
\begin{multline}
\label{eq40}
b_{j,g\leftrightarrow g',n}=\\
\begin{cases}
\displaystyle\int_{\min([\nu_{g+1/2},\nu_{g-1/2}]\cap[\nu_{g'+1/2},\nu_{g'-1/2}])}
^{\max([\nu_{g+1/2},\nu_{g-1/2}]\cap[\nu_{g'+1/2},\nu_{g'-1/2}])}b_{0,\nu_{0}}(T_{j,n})d\nu_{0}\;\;,\\
\;\;[\nu_{g+1/2},\nu_{g-1/2}]\cap[\nu_{g'+1/2},\nu_{g'-1/2}]\not=\emptyset\\
\\
0\;\;,\;\;[\nu_{g+1/2},\nu_{g-1/2}]\cap[\nu_{g'+1/2},\nu_{g'-1/2}]=\emptyset \;\;.
\end{cases}
\end{multline}
Equation~\eqref{eq39} presumes the radiation field near the cell boundary is well
approximated by a Planck function at the diffusive cell temperature.  With
$\sigma_{b(j,j'),g}$ on the right-hand-side of Eq.~\eqref{eq39}, it is clear that
changing $\sigma_{j\rightarrow j',g}\rightarrow\tilde{\sigma}_{j\rightarrow j',g}$ requires
a complimentary modification to the right-hand-side of our DDMC equation that balances
the field at boundaries~\citep{densmore2012}.
\begin{figure}
\begin{center}
\includegraphics[height=55mm]{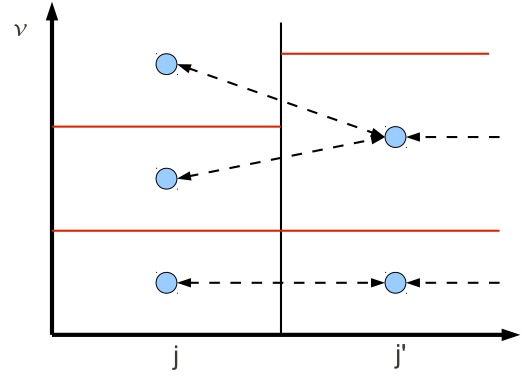}
\caption{Leakage probabilities can be thought of as weights
ascribed to edges on a set of topologically distinct graphs
in phase space.  The vertical and horizontal axes
represent frequency and spatial cell index, respectively.  The horizontal
in-cell partitions represent cell-specific frequency groupings.
The shaded circles represent possible cell-group locations a DDMC particle
or IMC particle may have and the dashed lines represent possible cell-group
leakages at cell interfaces.}
\label{fg1}
\end{center}
\end{figure}

In Fig.~\ref{fg1}, group bounds are given by horizontal lines at different values
along the vertical frequency axis and cells are enumerated along the horizontal
axis.  The circles are possible $(j,g)$ locations of IMC-DDMC particles and the
dashed lines represent transition possibilities for particles at those phase space
locations.  If all group bounds are aligned, the method reduces in complexity
to the standard multigroup approach and the set of $(j,g)$-transition graphs
become completely reducible.  If a leakage is sampled, the probability of a DDMC to
DDMC leakage transition is $(b_{j,g\leftrightarrow g_{D}',n}\sigma_{j\rightarrow j',g})
/(b_{j,g,n}\tilde{\sigma}_{j\rightarrow j',g})$ and the probability of an DDMC to IMC
leakage transition is $(b_{j,g\leftrightarrow g_{T}',n}\sigma_{b(j,j'),g})
/(b_{j,g,n}\tilde{\sigma}_{j\rightarrow j',g})$.  So the transition values for DDMC
are these probabilities that sum to 1 for each $(j,g)$ DDMC region.  If $(j,g)$ is
treated with IMC, then the group in the subsequent cell is determined by particle
frequency in the comoving frame of the boundary.

The fully hybridized comoving DDMC equation with elastic physical scattering is
\begin{multline}
\label{eq41}
\frac{1}{c}\frac{\partial\phi_{0,j,g}}{\partial t}+\bigg(
\sum_{j'}\tilde{\sigma}_{j\rightarrow j',g}+(1-\gamma_{j,g,n})(1-f_{j,n})\sigma_{a,j,g,n}\\
+f_{j,n}\sigma_{a,j,g,n}\bigg)\phi_{0,j,g}=f_{j,n}\gamma_{j,g,n}\sigma_{P,j,n}acT_{j,n}^{4}\\
+\frac{1}{V_{j}}\sum_{j'}V_{j'}\sum_{g_{D}'}
\frac{b_{j,g\leftrightarrow g_{D}',n}}{b_{j,g,n}}\sigma_{j'\rightarrow j,g_{D}'}\phi_{0,j,g_{D}'}\\
+\frac{1}{V_{j}}\sum_{j'}\sum_{g_{T}'}\int_{A_{b(j,j')}}\int_{\hat{\Omega}_{0}\cdot\vec{n}<0}
\int_{g\leftrightarrow g_{T}'}\ldots\\
P_{b(j,j')}(|\hat{\Omega}_{0}\cdot\vec{n}|)|\hat{\Omega}_{0}\cdot
\vec{n}|I_{0,\nu_{0}}d\nu_{0}d\Omega_{0}d^{2}\vec{r}\\
+\frac{\gamma_{j,g,n}(1-f_{j,n})}{V_{j}}\ldots\\
\sum_{g_{T}}\int_{V_{j}}\int_{4\pi}\int_{\nu_{g_{T}+1/2}}^{\nu_{g_{T}-1/2}}
\sigma_{a,j,n}I_{0,\nu_{0}}d\nu_{0}d\Omega_{0}d^{3}\vec{r}\\
+\gamma_{j,g,n}(1-f_{j,n})\sum_{g_{D}}\sigma_{a,j,g_{D},n}\phi_{0,j,g_{D}}
\end{multline}
where
\begin{multline}
\label{eq42}
\int_{g\leftrightarrow g'}f(\nu)d\nu=\\
\begin{cases}
\displaystyle\int_{\min([\nu_{g+1/2},\nu_{g-1/2}]\cap[\nu_{g'+1/2},\nu_{g'-1/2}])}
^{\max([\nu_{g+1/2},\nu_{g-1/2}]\cap[\nu_{g'+1/2},\nu_{g'-1/2}])}f(\nu)d\nu \;\;,\\
\;\;[\nu_{g+1/2},\nu_{g-1/2}]\cap[\nu_{g'+1/2},\nu_{g'-1/2}]\not=\emptyset\\
\\
0\;\;,\;\;[\nu_{g+1/2},\nu_{g-1/2}]\cap[\nu_{g'+1/2},\nu_{g'-1/2}]=\emptyset\;\;.
\end{cases}
\end{multline}
Groups represented by sub-indices $g_{D}$ or $g_{T}$ can be thought of as the complimentary
sets over frequency that are used during the advance of particles.  In other words, an
optimization over frequency may be applied such that the comoving group structure in a
time step is distinct from the user-prescribed group structure.  This re-grouping must not
degrade the accuracy of the observables of interest (for instance, a spectral tally
of particles leaving the domain).  Densmore's approach to multifrequency DDMC lumps adjacent
groups into a grey DDMC group~\citep{densmore2012}.  Indices $(j,g)$ and $(j,g+1)$ are combined
if their opacities multiplied by a characteristic cell length are greater than a threshold
number of mean free paths per cell~\citep{densmore2012,abdikamalov2012}.  Additionally, this
combined group has distinct Planck and Rosseland opacities,
\begin{equation}
\label{eq43}
\sigma_{P,j,g\cup g+1,n}=\frac{b_{j,g,n}\sigma_{a,j,g,n}+b_{j,g+1,n}\sigma_{a,j,g+1,n}}
{b_{j,g,n}+b_{j,g+1,n}} \;\;,
\end{equation}
and
\begin{equation}
\label{eq44}
\frac{1}{\sigma_{R,j,g\cup g+1,n}}=\frac{b_{j,g,n}/\sigma_{a,j,g,n}+b_{j,g+1,n}/\sigma_{a,j,g+1,n}}
{b_{j,g,n}+b_{j,g+1,n}} \;\;,
\end{equation}
respectively, where $g\cup g+1$ denotes a DDMC group union.  Equations~\eqref{eq43} and
\eqref{eq44} can be used in place of $\sigma_{a,j,g,n}$ in the formulas for leakage opacity
and in absorption sampling.  The increase in efficiency is obtained from
\begin{equation}
\label{eq45}
\gamma_{j,g\cup g+1,n}=\gamma_{j,g,n}+\gamma_{j,g+1,n} \;\;.
\end{equation}
For either $g$ or $g'$, $\gamma_{j,g\cup g+1,n}$ used in the DDMC effective out-scattering
coefficient, $(1-f_{j,n})(1-\gamma_{j,g\cup g+1,n})$, in place of the values on the left-hand-side of
Eq.~\eqref{eq45}.  For Densmore's simulations~\citep{densmore2012}, only opacities that are monotonic
in frequency are considered, so the group lumping method is conflated with a threshold frequency.
Nothing in Densmore's asymptotic analysis suggests that group lumping cannot be employed in frequency
regions away from $\nu=0$.  Densmore indeed notes the possibility of this extension~\citep{densmore2012}.

\section{Code Verifications}
\label{sec:CodeVer}

In the numerical verification results that follow, we do not apply group lumping or irreducible group graphs
(see Fig.~\ref{fg1}), because applying these generalizations should not significantly reduce computation time,
given the simple high-contrast opacity structures we use.  For the plot
legends, ``HMC'' stands for hybrid Monte Carlo and denotes instances of the method where both DDMC
and IMC play a significant role.  For simulations involving all or mostly DDMC, the plot label is
``DDMC''.  Otherwise, ``IMC'' is the label applied to computations that are transport dominant
or constrained to only use IMC (referred to as pure IMC in captions).

\subsection{Static Grid Verification}
\label{sec:Static}

Our first verification is in a static material with a multifrequency structure that
is amenable to analytic solution.  Specifically, we use the~\cite{su1999} picket fence opacity
structure.  We apply this picket fence opacity distribution to the static P$_{1}$ equations with
thermal coupling.  \cite{mcclarren2008} have solved the thermally coupled P$_{1}$ equations
in grey materials for several one-dimensional geometries.  The P$_{1}$ method
is higher order than diffusion, but still approximate.
This verification is therefore performed in an optically thick material where DDMC,
IMC, and P$_{1}$ should show quantitative agreement for a range of spatial grids.

The picket fence opacity dependence on frequency can be constructed as the limit of a
discrete multigroup distribution.  Partitioning the frequency grid into regular $\Delta\nu$
intervals, a portion $p_{1}$ of $\Delta\nu$ is attributed an opacity $\sigma_{1}$ while the
remainder $p_{2}=1-p_{1}$ is attributed an opacity $\sigma_{2}$.  The limit as $\Delta\nu\rightarrow 0$
of this alternating grouping is the picket fence opacity $\sigma(\nu)$.  Each picket at
some $\nu$ is dense over the real number line of $\nu$; meaning both picket values are in any
nonempty, open interval over the real number line for $\nu$.

The picket-fence distribution has the nice property that integrals of $I_{\nu}$ and $B_{\nu}$
over the dense groupings simplify when these integrands are assumed to be smooth~\citep{su1999}.
Su and Olson solve the transport equations with the picket fence opacity to obtain a semi-analytic
result.  For specific values of $\sigma_{1}$, $p_{1}$, $\sigma_{2}$, and $p_{2}$, we may
develop a simple generalization of McClarren's P$_{1}$ solution that includes a rudimentary test
of multifrequency for {\tt SuperNu}.

Neglecting scattering, taking the zeroth and first angular moments of Eq.~\eqref{eq1}, and
integrating the result over the set of frequencies that only yield a contribution from picket
$g\in\{1,2\}$, the thermal picket fence P$_{1}$ equations in planar 1D geometry are
\begin{equation}
\label{eq46}
\frac{1}{c}\left(\frac{\partial E_{g}}{\partial t}+\frac{\partial F_{g}}{\partial z}\right)
=\sigma_{g}(p_{g}aT^{4}-E_{g})+p_{g}S \;\;,
\end{equation}
\begin{equation}
\label{eq47}
\frac{1}{c}\frac{\partial F_{g}}{\partial t}+\frac{c}{3}\frac{\partial E_{g}}{\partial z}
=-\sigma_{g}F_{g} \;\;,
\end{equation}
\begin{equation}
\label{eq48}
\frac{C_{v}(T)}{c}\frac{\partial T}{\partial t}=\sum_{g'=1}^{2}\sigma_{g}E_{g}-\bar{\sigma}
aT^{4}
\end{equation}
where $z$ is the spatial coordinate, $E_{g}=\phi_{g}/c$ is the radiation energy density,
$\bar{\sigma}=p_{1}\sigma_{1}+p_{2}\sigma_{2}$ and $S$ is an external source.  The system
of equations is linearized with $C_{v}(T)= a'T^{3}$~\citep{mcclarren2008,su1999}.  Furthermore,
we apply the usual non-dimensionalizations for convenience~\citep{mcclarren2008,su1999}:
\begin{equation}
\label{eq49}
x=\bar{\sigma}z\;\;,\;\;\epsilon=\frac{4a}{a'}\;\;,\;\;\tau=\epsilon c\bar{\sigma}t\;\;,\;\;
w_{g}=\frac{\sigma_{g}}{\bar{\sigma}}
\end{equation}
and
\begin{equation}
\label{eq50}
\mathcal{E}_{g}=\frac{E_{g}}{aT_{r}^{4}}\;\;,\;\;\mathcal{F}_{g}=\frac{F_{g}}{aT_{r}^{4}}\;\;,
\;\;\mathcal{M}=\frac{T^{4}}{T_{r}^{4}}\;\;,\;\;Q=\frac{S}{\bar{\sigma}aT_{r}^{4}}
\end{equation}
where $T_{r}$ is a reference temperature.  Incorporating Eqs.~\eqref{eq49} and~\eqref{eq50},
Eqs.~\eqref{eq46},~\eqref{eq47} and~\eqref{eq48} become
\begin{equation}
\label{eq51}
\epsilon\frac{\partial\mathcal{E}_{g}}{\partial\tau}+\frac{1}{c}
\frac{\partial\mathcal{F}_{g}}{\partial x}=w_{g}(p_{g}\mathcal{M}-\mathcal{E}_{g})+p_{g}Q
\;\;,
\end{equation}
\begin{equation}
\label{eq52}
\epsilon\frac{\partial\mathcal{F}_{g}}{\partial\tau}+\frac{c}{3}\frac{\partial\mathcal{E}_{g}}
{\partial x}=-w_{g}\mathcal{F}_{g} \;\;,
\end{equation}
\begin{equation}
\label{eq53}
\frac{\partial\mathcal{M}}{\partial\tau}=\sum_{g'=1}^{2}w_{g}\mathcal{E}_{g}-\mathcal{M}\;\;.
\end{equation}

If Eqs.~\eqref{eq51},~\eqref{eq52}, and~\eqref{eq53} are Laplace transformed over time and
reduced to equations for only energy density, what results are two fourth order linear
differential equations in space with coefficients that are algebraically irrational functions
of the Laplace variable, $s$.  Denoting transformed quantities with a tilde, the equations
can be solved to obtain $\tilde{\mathcal{E}}_{g}(x,s)$ and subsequently inverse Laplace-transformed
to $\mathcal{E}_{g}$.  The inverse Laplace transform of $\tilde{\mathcal{E}}_{g}$ is not known analytically,
but it may be performed numerically.  The material temperature is then proportional to a
temporal convolution of the opacity weighted sum of the radiation energy densities,
$\mathcal{E}_{g}$, and $e^{-\tau}$~\citep{mcclarren2008}.

To leverage the work done by McClarren, we constrain the $g=1$ picket with $w_{1}/w_{2}\ll 1$
or $w_{1}\approx 0$.  The $g=2$ picket opacity is constrained to $w_{2}=\epsilon$.  The relation
between the non-dimensional optically thick picket and the non-dimensional heat capacity does not
have physical justification but is merely a device to make certain expressions Laplace invertible.
With $p_{2}w_{2}\approx 1$, $p_{2}\approx 1/\epsilon$ and $p_{1}\approx 1-1/\epsilon$, the value
of $p_{1}$ can be comparable to $p_{2}$ despite the disparity in opacity strength.  To our knowledge,
the approach of relating $w_{2}$ with $\epsilon$ is novel.

With the above constraints and $Q=\delta(x)\delta(\tau)$, the Fourier-Laplace transformed
system of equations is
\begin{equation}
\label{eq54}
(k^{2}+3\epsilon^{2}s^{2})\tilde{\tilde{\mathcal{E}}}_{1}=3\epsilon s
\left(1-\frac{1}{\epsilon}\right) \;\;,
\end{equation}
\begin{equation}
\label{eq55}
(k^{2}+3\epsilon^{2}s(s+2))\tilde{\tilde{\mathcal{E}}}_{2}=3(s+1) \;\;,
\end{equation}
\begin{equation}
\label{eq56}
(s+1)\tilde{\tilde{\mathcal{M}}}\approx w_{2}\tilde{\tilde{\mathcal{E}}}_{2} \;\;,
\end{equation}
where double tilde indicates a Fourier-Laplace transformed quantity.  The remainder of
the derivation follows closely from~\cite{mcclarren2008} and is not repeated here.  Solving the
above equations yields the following kernel planar solutions for the constrained picket fence
distribution:
\begin{equation}
\label{eq57}
\mathcal{E}_{1}=\left(1-\frac{1}{\epsilon}\right)\frac{\sqrt{3}}{2}\delta(\tau-\epsilon\sqrt{3}|x|)
\;\;,
\end{equation}
\begin{multline}
\label{eq58}
\mathcal{E}_{2}=\frac{1}{\epsilon}\frac{\sqrt{3}}{2}e^{-\tau}
\Bigg[\frac{\tau I_{1}(\sqrt{\tau^{2}-3\epsilon^{2}x^{2}})}
{\sqrt{\tau^{2}-3\epsilon^{2}x^{2}}}\Theta(\tau-\epsilon\sqrt{3}|x|)+\\
I_{0}(\sqrt{\tau^{2}-3\epsilon^{2}x^{2}})\delta(\tau-\epsilon\sqrt{3}|x|)\Bigg] \;\;,
\end{multline}
and
\begin{equation}
\label{eq59}
\mathcal{M}=
\frac{\sqrt{3}}{2}e^{-\tau}I_{0}(\sqrt{\tau^{2}-3\epsilon^{2}x^{2}})\Theta(\tau-\epsilon\sqrt{3}|x|)
\end{equation}
where $\Theta$ is the Heaviside function, and $I_{0}$ ($I_{1}$) is the 0-order (1-order) modified
Bessel function.  Setting $\epsilon=1$ yields McClarren's result~\citep{mcclarren2008}.  The properties
of the P$_{1}$ equations allow for the application of a simple planar to spherical Green's function mapping
\citep{mcclarren2008},
\begin{equation}
\label{eq60}
G_{\text{point}}(r,\tau)=-\frac{1}{2\pi r}\left.\frac{\partial G_{\text{plane}}}{\partial x}\right|_{x=r}
\;\;,\;\;r>0 \;\;,
\end{equation}
where $G_{\text{point}}$ and $G_{\text{plane}}$ are the Green functions for 1D spherically symmetric
and 1D planar geometry, respectively.  Since Eqs.~\eqref{eq57},~\eqref{eq58}, and~\eqref{eq59} were
derived with a kernel source, $\mathcal{E}_{g}$ or $\mathcal{M}$ may be substituted into the right
hand side of Eq.~\eqref{eq60}.  The unitless spherically symmetric material temperature kernel is
\begin{multline}
\label{eq61}
\mathcal{M}(r,\tau)=\\\frac{3\sqrt{3}}{4\pi}e^{-\tau}\epsilon^{2}
\frac{I_{1}(\sqrt{\tau^{2}-3\epsilon^{2}r^{2}})}{\sqrt{\tau^{2}-3\epsilon^{2}r^{2}}}
\Theta(\tau-\epsilon\sqrt{3}r)
+\\
\frac{3}{4\pi}e^{-\tau}I_{0}(\sqrt{\tau^{2}-3\epsilon^{2}r^{2}})\epsilon
\frac{\delta(\tau-\epsilon\sqrt{3}r)}{r} \;\;.
\end{multline}

\begin{figure}
\includegraphics[height=70mm]{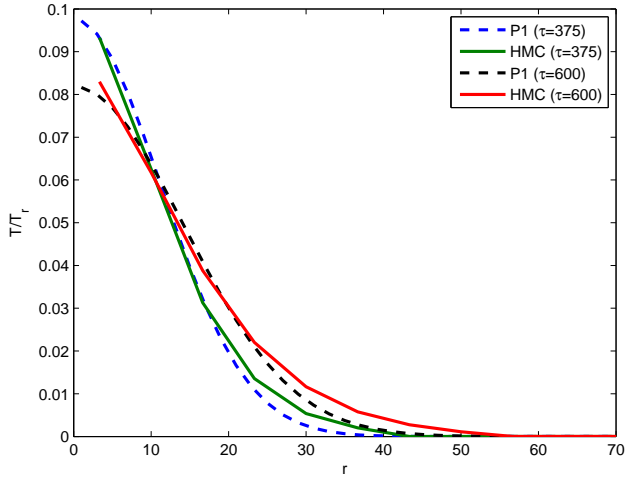}
\caption{
P$_{1}$ (dashed) and IMC-DDMC (solid) unitless material temperature profiles at two
different (mean free) times plotted as a function of unitless radius (mean free paths).
These curves are solutions to the picket-fence opacity problem described in Section~\ref{sec:Static}.
At 6.67 mean free paths per spatial cell, only DDMC is applied to radiation interacting with the
$g=2$ picket.  For the $g=1$ picket, IMC allows the radiation to stream out of the domain and
hence plays no role in the thermal state.
The combined MC particle fields produce an accurate solution tally with respect to
the analytic P$_{1}$ solution at coarse spatial resolution.
}
\label{fg2}
\end{figure}

\begin{figure}
\includegraphics[height=70mm]{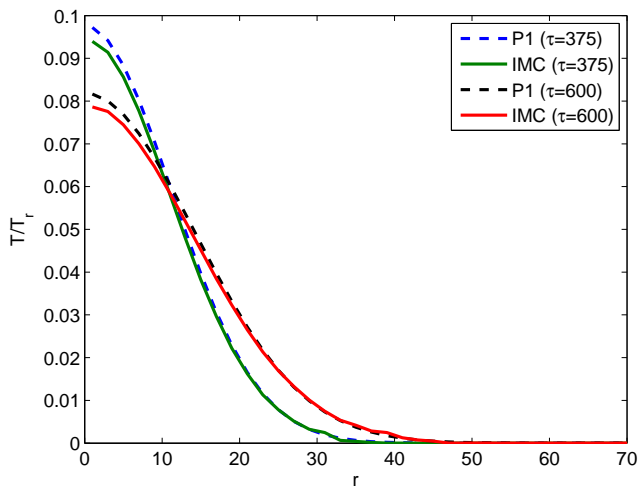}
\caption{
P$_{1}$ (dashed) and IMC-DDMC (solid) unitless material temperature profiles at two
different (mean free) times plotted as a function of unitless radius (mean free paths).
These curves are solutions to the picket-fence opacity problem described in Section~\ref{sec:Static}.
At 2 mean free paths per spatial cell, only IMC is applied to both $g=1$ and $g=2$ radiation
fields.  Relative to the coarse grid solutions of Fig.~\eqref{fg2}, the IMC-DDMC solution tally
is in closer agreement with the P$_{1}$ solutions.
}
\label{fg3}
\end{figure}

The picket fence opacity is simple to implement in IMC-DDMC.  Giving the nature of our picket fence
constraint, we do not need group lumping or non-uniform particle transition probabilities.
In this case, $\sigma_{P,j,n}=\bar{\sigma}$, $\gamma_{j,g,n}=p_{g}\sigma_{g}/\bar{\sigma}=p_{g}w_{g}$,
and $\sigma_{a,j,g,n}=\sigma_{P,j,g,n}=\sigma_{R,j,g,n}$.  We experiment with $\epsilon=2$, 15 or 50
cells over a spherical domain of 100 mean free paths, 400,000 source particles per time step over 100
time steps from $\tau=300$ to $\tau=600$.  An MC particle propagates with DDMC when
there are greater than 3 mean free paths per the particle's $(j,g)$ coordinate.  In order to simulate a
problem with an instantaneous point source, we initialize the material temperature using the analytic 
solution and instantiate the initial MC particle field accordingly.  Figure~\ref{fg2} contains 15 cell
IMC-DDMC material temperature data plotted against Eq.~\eqref{eq61}.
The 50 cell case should be more converged at each time relative to the results in Fig.~\ref{fg2}.
Indeed, in Fig.~\ref{fg3} we see that IMC is now applied everywhere in the spectrum and the solution
is more converged.

We next measure the error of IMC and DDMC relative to the P$_{1}$ temperature solution
for different numbers of spatial cells.
A grid convergence study for a stochastic method must have enough particles per cell
to make the statistical error negligible.  Another consideration involves a peculiarity
of IMC specifically.
Much work in IMC pertaining to the effect of the spatial grid on the continuous
transport has been to formally characterize a pathology referred to as
teleportation error~\citep{mckinley2003,densmore2011}.  Teleportation error occurs
in IMC when there are many absorption mean free paths per cell but not many absorption
mean free times per time step.  Particle energies absorbed at one location in a cell
may be re-emitted in a subsequent time step many mean free paths away
from the absorption locations~\citep{mckinley2003}.  \cite{densmore2011} demonstrates
formally that a piecewise constant representation of scalar flux along with a time
step that resolves a mean free time does not asymptotically converge to a correct
discretization of the diffusion equation.

\begin{figure}
\includegraphics[height=70mm]{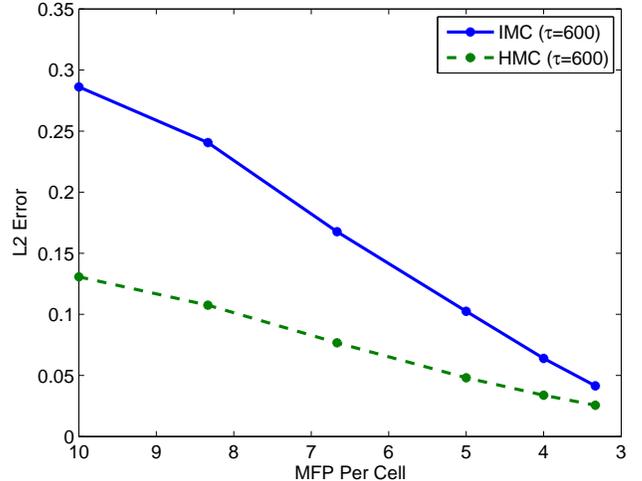}
\caption{
IMC (solid) and IMC-DDMC (dashed) L$_{2}$ temperature error relative to the P$_{1}$
solution plotted over the number of mean free paths per cell at 600 mean free times.
From 10 to 30 cells (10 to about 3.33 mean free paths per cell), IMC and DDMC appear to 
linearly converge.  The error from IMC-DDMC (where DDMC contributes non-negligibly to
the temperature) is found to be distinctively lower than that of pure IMC down to
3.33 mean free paths per cell.}
\label{fg44}
\end{figure}

Despite these complications, there does exist literature to indicate that the IMC
temperature error scales with $\Delta r$ where $\Delta r$ is a typical cell size of 
the simulation (if not the actual cell length for a one dimensional simulation).
In investigating a method to emit particles at sub-cell deposition locations,
\cite{irving2011} measure total relative error of standard IMC for a 1D Marshak
wave problem in planar geometry to a converged solution
at several temporal and spatial resolutions.  While not explicitly stated, their
findings appear to yield an total relative error of about $9/J$, particularly between
$J=10$ and $J=30$ cells, for several time step sizes.  \cite{cheatham2010} plots 
relative errors of IMC with respect to a grey form of the \cite{su1999} solutions in
two cells of a simulation.  This IMC error also appears to roughly
scale linearly with cell width for each cell.  Having set each simulation to have well
over 13,000 particles generated per cell per time step and testing in a regime where
IMC teleportation should be minimal, Fig.~\ref{fg44} indicates our code indeed achieves
an approximately linear scaling in L$_{2}$ relative error for both IMC and DDMC as
well.

Our results indicate that DDMC computes temperature with higher fidelity at low
cell resolution for this particular problem.
Having obtained good agreement with a static grid analytic solution, we present
and test a manufactured solution the next section that allows for outflow and a group
structure that includes scattering and absorption.

\subsection{Manufactured Verification}
\label{sec:Manu}

The Method of Manufactured Solutions (MMS)~\citep[p.~219]{oberkampf2010} provides an avenue of
code verification that is useful for problems with governing equations that are not amenable
to a direct solution with a prescribed source.  As the name MMS suggests, one postulates, or 
manufactures, a solution~\citep{mcclarren2008b,warsa2010}.  The next step simply involves 
incorporating this postulated answer into the system of equations to see what additional terms 
are produced through calculus and algebra~\citep{mcclarren2008b}.  The additional terms can be 
included in an appropriate code as artificial sources.  If the routines in the code function and 
interface correctly, the numerical experiment should reproduce the manufactured solution.
Manufactured solutions have been developed by~\cite{warsa2010} for discrete ordinate codes
modeling diffusive problems and by~\cite{mcclarren2008b} for
planar, grey radiation-hydrodynamics problems in the optically thick and optically thin limits.
We draw from these solutions as well as analytic forms presented by~\cite[p.~474]{mihalas1984},
and by~\cite{pinto2000} for high velocity outflow problems.

Our manufactured solution has two groups and the temperature and radiation fields are constant
in time and over the space of the expansion.  As will be shown, in order to achieve an ostensibly
simple solution, the code must have a time dependent source that can be distinct in
form for each group.  The source must supply energy to counteract the non-trivial adiabatic cooling
of the field.  The higher energy group has pure elastic scattering and the remaining group has both
elastic scattering and absorption.  These groups must couple through Doppler shifting in a way that
appropriately balances the supply of energy to each group.  Moreover, the solutions must exhibit an
invariance with respect to several wavelength or frequency grid values.

The constrained material properties are
\begin{subequations}
\label{eq62}
\begin{align}
& \vec{U}(r,t)=\frac{U_{\max}\vec{r}}{R(t)}\;\;, \\
& \rho(r,t)=\rho(t)=\frac{3M}{4\pi R(t)^{3}}\;\;,
\end{align}
\end{subequations}
where $\vec{U}(r,t)$ is fluid velocity, $\rho(r,t)$ is density, $U_{\max}$ is the maximum outflow
speed, $R(t)=R(0)+U_{\max}t$ is the outer expansion radius, $\vec{r}$ is the Eulerian
position vector, and $M$ is the total outflow mass.  The frequency integrated manufactured
radiation intensity and temperature are
\begin{subequations}
\label{eq63}
\begin{align}
& I_{0}(r,t) = \frac{\phi_{m}}{4\pi} \;\;,\\
& T(r,t) = T_{m}=(\phi_{m}/(ca))^{1/4}\;\;,
\end{align}
\end{subequations}
respectively.  The subscript $m$ denotes a manufactured value.  For a general
frequency grid, $\nu_{G+1/2}<\nu_{G-1/2}<\ldots <\nu_{1/2}$, and group index,
$g\in\{1\ldots G\}$, we constrain the absorption opacity to be
pre-grouped as
\begin{equation}
\label{eq64}
\sigma_{0,\nu_{0},a} = \sigma_{g}(t) = \kappa_{g}\rho(t) \;\;,
\end{equation}
where $\kappa_{g}$ is constant within group $g$.  The differential
scattering opacity has the form
\begin{equation}
\label{eq65}
\sigma_{0,s}(\nu_{0}'\rightarrow\nu_{0},\hat{\Omega}_{0}'\cdot\hat{\Omega}_{0})
=\frac{\sigma_{s}(t)}{4\pi}\delta(\nu_{0}'-\nu_{0})=
\rho(t)\frac{\kappa_{s}}{4\pi}\delta(\nu_{0}'-\nu_{0})\;\;,
\end{equation}
where $\kappa_{s}$ is a constant.
We now may express the manufactured multifrequency solution as an opacity-dependent 
superposition of the normalized Planck function and a piecewise constant function,
\begin{multline} 
\label{eq66}
\varphi_{m,\nu_{0}}=\left(\frac{\sigma_{s}}{\sigma_{g}+\sigma_{s}}\right)
\frac{\varphi_{m,s,g}}{\Delta\nu_{g}}+
\left(\frac{\sigma_{g}}{\sigma_{g}+\sigma_{s}}\right)b_{\nu_{0}}\;\;,\\
\nu_{0}\in[\nu_{g+1/2},\nu_{g-1/2}]\;\;,
\end{multline}
where $I_{0,\nu_{0}}=I_{0}\varphi_{m,\nu_{0}}$, $\varphi_{m,s,g}$ is the uniform non-thermal
contribution to $g$ and $b_{\nu_{0}}$ is the normalized Planck function.  With particular
choices of opacity, frequency grid, and $\varphi_{m,s,g}$, the integral of $\varphi_{\nu_{0}}$
may be constrained to 1.

Incorporating the form of the opacities into the transport and temperature
equations, the system to be solved with manufactured sources is
\begin{multline}
\label{eq67}
\left(1+\hat{\Omega}_{0}\cdot\frac{\vec{U}}{c}\right)
\frac{1}{c}\frac{D I_{0,\nu_{0}}}{Dt}+
\hat{\Omega}_{0}\cdot\nabla I_{0,\nu_{0}}-
\frac{1}{c}\hat{\Omega}_{0}\cdot\nabla\vec{U}\cdot\hat{\Omega}_{0}\nu_{0}
\frac{\partial I_{0,\nu_{0}}}{\partial\nu_{0}}\\
-\frac{1}{c}\hat{\Omega}_{0}\cdot\nabla\vec{U}\cdot
(\mathbf{I}-\hat{\Omega}_{0}\hat{\Omega}_{0})\cdot\nabla_{\hat{\Omega}_{0}}I_{0,\nu_{0}}+\\
\frac{3}{c}\hat{\Omega}_{0}\cdot\nabla\vec{U}\cdot\hat{\Omega}_{0}I_{0,\nu_{0}}=
\frac{1}{4\pi}\sigma_{g}b_{\nu_{0}}(T)acT^{4}-(\sigma_{g}+\sigma_{s})I_{0,\nu_{0}}\\
+\frac{\sigma_{s}}{4\pi}\int_{4\pi}I_{0,\nu_{0}}(\vec{r},\hat{\Omega}_{0}',t)d\Omega_{0}'
+ \frac{S_{m,\phi,\nu_{0}}(r,t)}{4\pi}\;\;,
\end{multline}
and
\begin{multline}
\label{eq68}
C_{v}\frac{DT}{Dt}=\\
\sum_{g=1}^{G}\int_{4\pi}\int_{\nu_{g+1/2}}^{\nu_{g-1/2}}\sigma_{g}
(I_{0,\nu_{0}}-ac T^{4}b_{0,\nu_{0}})d\nu_{0}d\Omega_{0}\\
+S_{m,T}(r,t)\;\;,
\end{multline}
where $S_{m,\phi,\nu_{0}}$ and $S_{m,T}$ are the manufactured radiation and material
sources to be determined.  At implementation, $S_{m,T}$ can be treated with
the usual Fleck factor re-balance of material source terms.  With the manufactured
solutions specified, the source terms are found to be
\begin{multline}
\label{eq69}
\frac{S_{m,\phi,\nu_{0}}}{\phi_{m}}=
\frac{U_{\max}}{cR(t)}\left(3\varphi_{m,\nu_{0}}-
\nu_{0}\frac{\partial\varphi_{m,\nu_{0}}}{\partial\nu_{0}}\right)\\
+\sigma_{g}(\varphi_{m,\nu_{0}}-b_{\nu_{0}}(T_{m}))\;\;,
\end{multline}
and
\begin{equation}
\label{eq70}
\frac{S_{m,T}}{\phi_{m}}=\sum_{g=1}^{G}\sigma_{g}(b_{g}(T_{m})-\varphi_{m,g})
\;\;,
\end{equation}
where $\varphi_{m,g}=\int_{\nu_{g+1/2}}^{\nu_{g-1/2}}\varphi_{m,\nu_{0}}d\nu_{0}$.
Integration of Eq.~\eqref{eq69} over group $g$ yields
\begin{multline}
\label{eq71}
\frac{S_{m,\phi,g}}{\phi_{m}}=\\
\frac{U_{\max}}{cR(t)}\left(4\varphi_{m,g}-\nu_{g-1/2}\varphi_{m,\nu_{g-1/2}}+
\nu_{g+1/2}\varphi_{m,\nu_{g+1/2}}\right)\\
+\sigma_{g}(\varphi_{m,g}-b_{g}(T_{m}))
\end{multline}
We now construct a two group instantiation of Eqs.~\eqref{eq70} and
\eqref{eq71} that is simple to implement but is a good test of energy balance
and group coupling in the code.  A scattering opacity $\sigma_{s}=0.1\rho(t)$
is applied along with absorption opacities $\sigma_{1}=0$ and
$\sigma_{2}=0.1\rho(t)$.  The sub-group profile of the radiation energy
density in $g=1$ is constant.  Thus, particle frequency may be sampled
uniformly in $g=1$.  If the group domains are implemented in wavelength,
then the MC source particle wavelengths in $g=1$ must be calculated as the
reciprocal of a uniform sampling between the reciprocals of the group wavelength
bounds.  Since radiation only redshifts, the upwind group approximation~\citep[p.~475]
{mihalas1984} for the group-interface terms in Eq.~\eqref{eq71} must be a reasonable
approach for Eq.~\eqref{eq66} and the MC process described.  Equation~\eqref{eq71}
consequently may be expressed as
\begin{equation}
\label{eq72}
\frac{S_{m,\phi,1}}{\phi_{m}}=
\frac{U_{\max}}{cR(t)}\left(4\varphi_{m,1}+\nu_{3/2}\varphi_{m,\nu_{3/2}}^{+}\right)\;\;,
\end{equation}
and
\begin{multline}
\label{eq73}
\frac{S_{m,\phi,2}}{\phi_{m}}=
\frac{U_{\max}}{cR(t)}\left(4\varphi_{m,2}-\nu_{3/2}\varphi_{m,\nu_{3/2}}^{+}\right)+\\
\sigma_{2}(\varphi_{m,2}-b_{2}(T_{m}))\;\;,
\end{multline}
where the plus superscript denotes evaluation on the right side of the frequency bound.
We exploit our ability to choose a frequency grid that further simplifies the form of the
source terms.
Using Eq.~\eqref{eq66}, if $\nu_{5/2}=0$, $\varphi_{m,s,1}=\varphi_{m,s,2}=1/2$ and the integral
of $b_{\nu_{0}}$ over $\nu_{0}\in[0,\nu_{3/2}]$ is 1/2, then $\varphi_{2}=b_{2}=1/2$ and the
source terms become
\begin{equation}
\label{eq74}
\frac{S_{m,\phi,1}}{\phi_{m}}=
\frac{U_{\max}}{cR(t)}\left(2+\frac{1}{2}\frac{\nu_{3/2}}{\Delta\nu_{1}}\right)\;\;,
\end{equation}
\begin{equation}
\label{eq75}
\frac{S_{m,\phi,2}}{\phi_{m}}=
\frac{U_{\max}}{cR(t)}\left(2-\frac{1}{2}\frac{\nu_{3/2}}{\Delta\nu_{1}}\right)\;\;,
\end{equation}
and
\begin{equation}
\label{eq76}
S_{m,T} = 0
\end{equation}
Newton iteration yields $h\nu_{3/2}/kT_{m}\approx3.503$
to obtain $b_{2}(T_{m})=1/2$.  To satisfy Eq.~\eqref{eq76}, the MC process
must deposit the correct energy in $g=2$ directly from radiation generated in $g=2$ and
indirectly from redshifting radiation originating in $g=1$.  The strength of the group
coupling is quantified with $\nu_{3,2}/\Delta\nu_{1}$.

The numerical results for our manufactured solution include strong and weak Doppler coupling
for pure IMC, IMC-DDMC, and pure DDMC.  For the IMC-DDMC hybrid, IMC is employed in the
pure scattering group and DDMC is employed in the thermally coupled group.
We note there are several ways to implement the Doppler shifting in the pure DDMC test
that are equivalent.  For instance, the group bound terms on the right-hand-side of Eq.
\eqref{eq30} can be implemented with the upwind approximation as probabilities of redshift
during the DDMC process if the scattering is elastic.
For the following results, each DDMC particle has its wavelength or frequency sampled according
to the sub-group distribution (which is constant in this case) after transport.  The frequency
value is redshifted by the same formula that lowers the particle's energy weight.  If the
new value of frequency is located in a new group, the particle is transferred to that group
for the next time step.

For the first test, we set $\nu_{3/2}/\Delta\nu_{1}\approx 0.036$ which of course is small
relative to 4.  The other domain quantities are set in a manner that makes adiabatic cooling
of the trapped radiation field non-negligible: $U_{\max}=10^{9}$ cm/s, $R(0)=1.728\times10^{14}$
cm, $t\in [172,800,181,440]$ seconds (or 2 to 2.1 days), $M=10^{33}$ g, $C_{v}=2\times10^{7}\rho$,
$T_{m}=1.1602\times 10^{7}$ K, and a wavelength grid of $\{\lambda_{1/2},\lambda_{3/2},\lambda_{5/2}\}=
\{1.239\times10^{-9},3.542\times10^{-8},1.2398\times10^{-3}\}$ cm.

\begin{figure*}
\centering
\subfloat[]{\includegraphics[height=65mm]{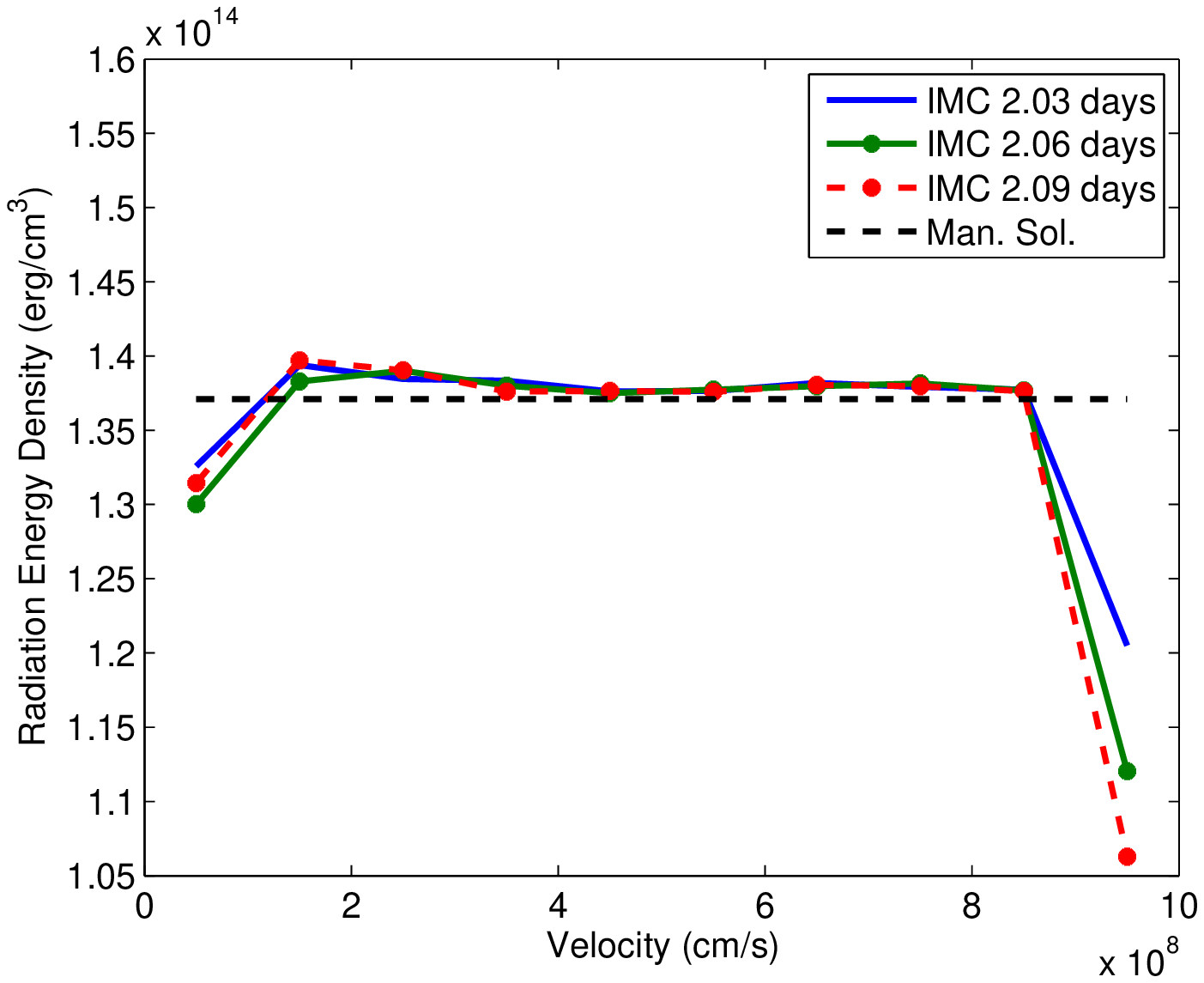}\label{fg5}}
\subfloat[]{\includegraphics[height=65mm]{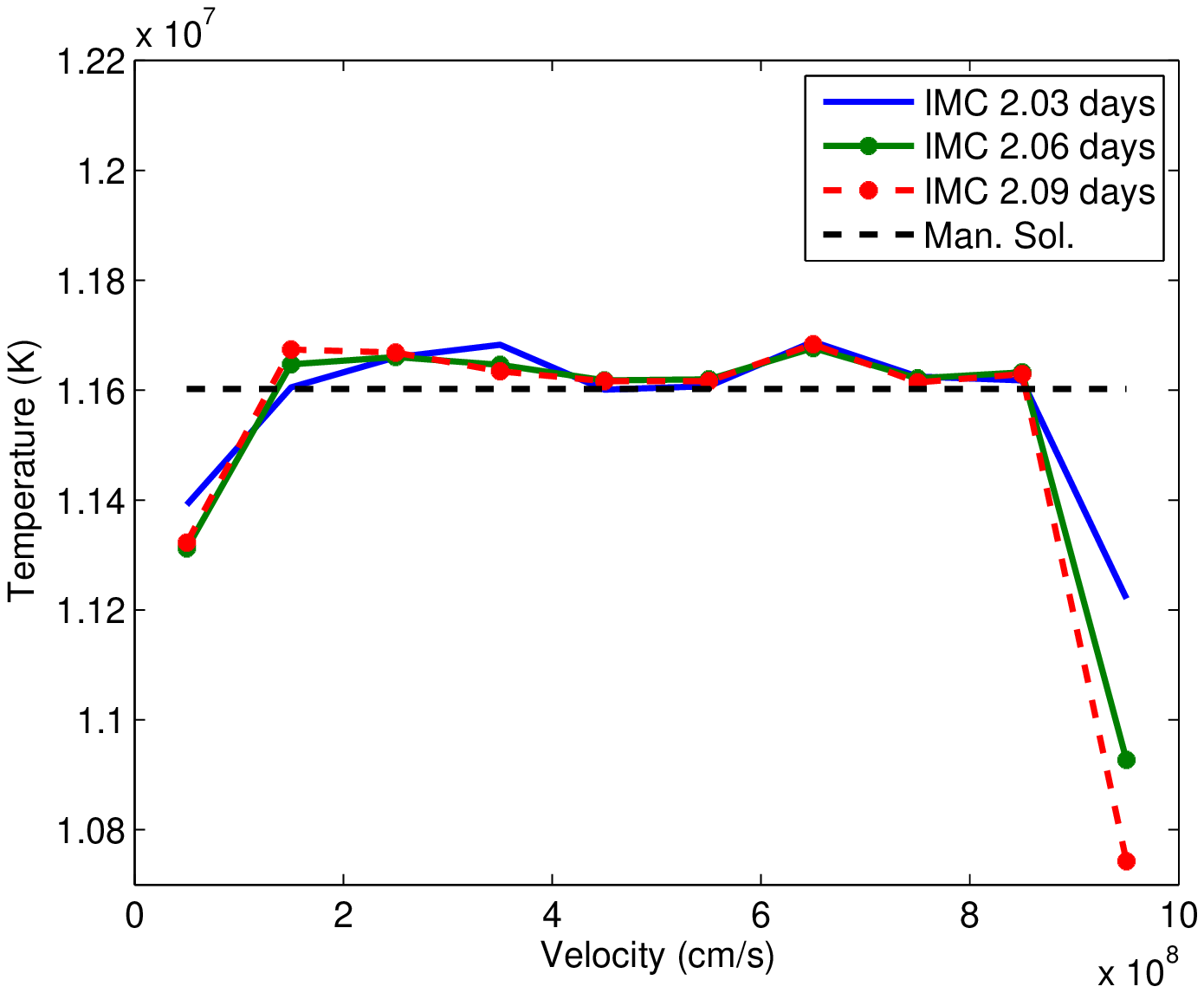}\label{fg6}}\\
\subfloat[]{\includegraphics[height=65mm]{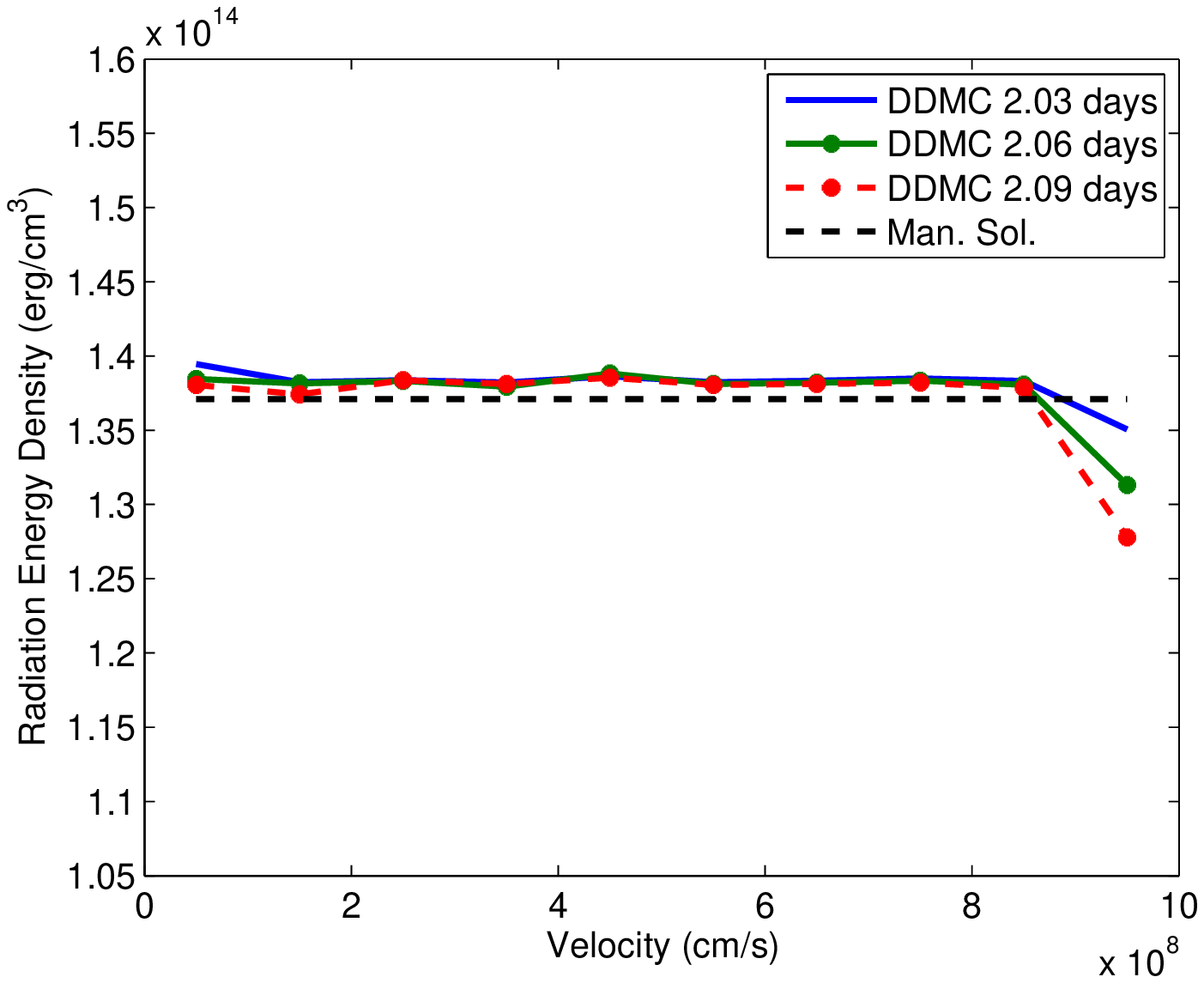}\label{fg7}}
\subfloat[]{\includegraphics[height=65mm]{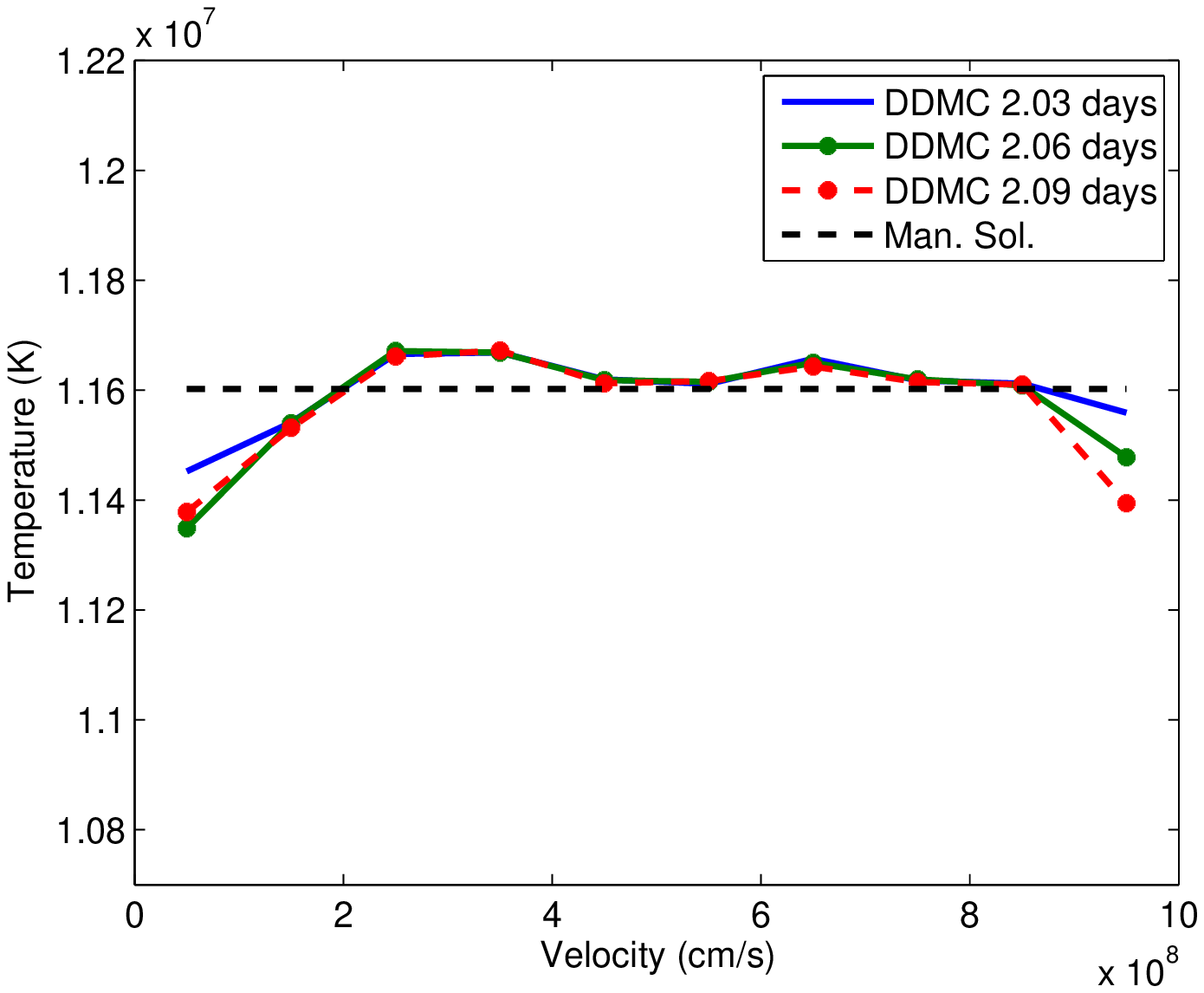}\label{fg8}}\\
\subfloat[]{\includegraphics[height=65mm]{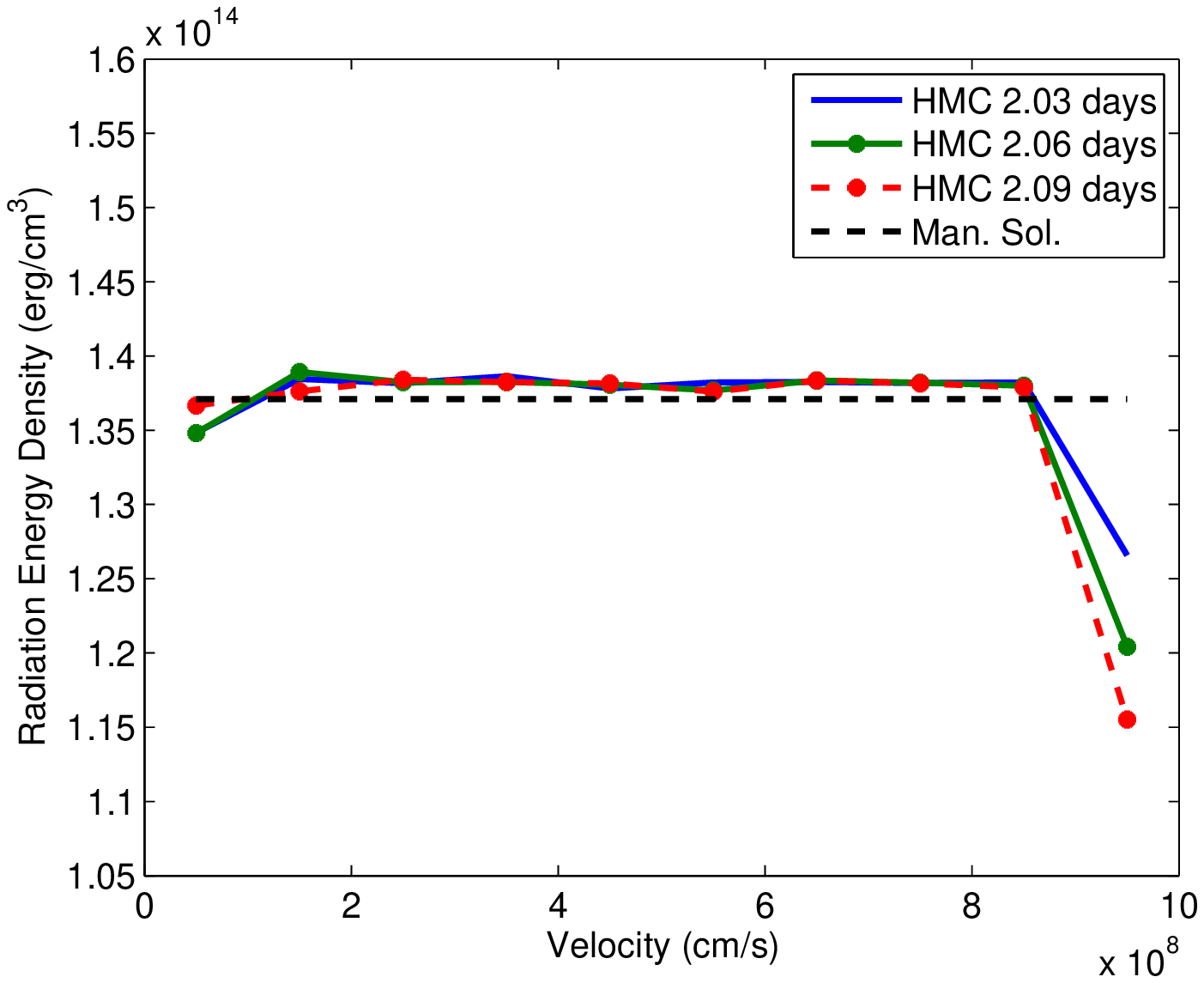}\label{fg9}}
\subfloat[]{\includegraphics[height=65mm]{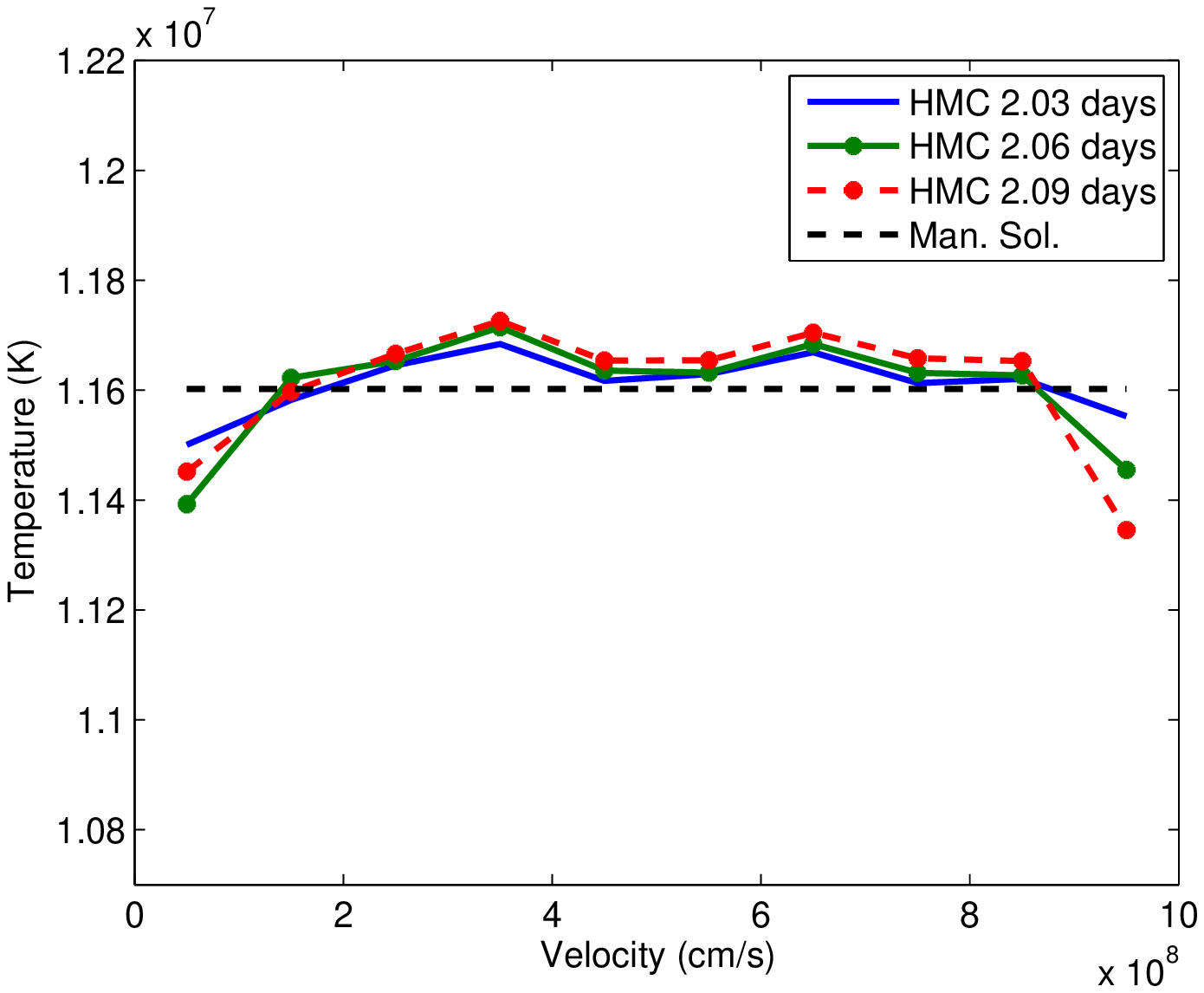}\label{fg10}}
\caption{
Manufactured (dashed) and simulated radiation energy densities (left figures) and temperatures
(right figures) at three different times (solid, dot-solid, and dot-dashed).  Figures~\ref{fg5}
and~\ref{fg6} have IMC, Figs.~\ref{fg7} and~\ref{fg8} have DDMC, and Figs.~\ref{fg9} and
\ref{fg10} have IMC-DDMC.  These MC profiles are generated by implementing the manufactured
source described in Section~\ref{sec:Manu}.  Specifically, data in Figs.~\ref{fg5},~\ref{fg6},
\ref{fg7}, and~\ref{fg8} use a wavelength grid that admits a weak Doppler coupling between
the groups while data in Figs.~\ref{fg9} and~\ref{fg10} use a wavelength grid that admits a
strong Doppler coupling between the groups.
The IMC deviation from uniformity towards the origin is apparently due to statistical noise
and can be reduced by increasing the number of source or initial particles allocated per cell.
The deviation at the outer bound is due to radiation escaping into an adjacent vacuum.
This test verifies that the code reproduces the analytic solution in each of the modes of
operation: IMC, DDMC and IMC-DDMC.
}
\end{figure*}

The pertinent computational quantities are: 10 time steps, $J=10$ spatial cells,
400,000 source particles generated per time step, 400,000 initial particles.  For the
specification provided, we expect the code to produce reasonable agreement to the manufactured
profiles.  Figures~\ref{fg5} and~\ref{fg6} have pure IMC data for radiation energy density and
temperature at several times, respectively.


Figures~\ref{fg7} and~\ref{fg8} have pure DDMC data for radiation energy density
and temperature at several times, respectively.  The DDMC results indicate the method
does not predict as much leakage at the outer cell as IMC through the observed times.
DDMC appears to produce less noise near the origin for the problem depicted.

We do not show the IMC-DDMC results for this problem here but note that
they too reproduce the manufactured profiles over the computed time scale.  With the
computation time of pure DDMC scaled to 1, the following table has the relative
computation times of IMC-DDMC and IMC for the weak coupling manufactured source test.
\begin{table}[H]
\caption{Weak Group Coupling Computation Times}
\centering
\resizebox{40 mm}{!}{
\begin{tabular}{|l|r|}
\hline
Method & Scaled Time \\
\hline
DDMC & 1 \\
\hline
HMC & 27.25\\
\hline
IMC & 147.39\\
\hline
\end{tabular}
}
\end{table}

We now change the coupling term to be comparable to 4 with a value of 
$\nu_{3/2}/\Delta\nu_{1}\approx 3.0$.
This is a more strenuous test on the code's ability to tally the correct redshift rates
because the manufactured source in $g=2$ is much reduced.  The modified wavelength grid
has $\lambda_{1/2}=2.656838745\times10^{-8}$ cm to implement this coupling strength.
Otherwise, all quantities are the same from the weak coupling test.
The IMC-DDMC radiation energy density and temperature results for this test are shown in 
Fig.~\ref{fg9} and~\ref{fg10}.  Results for pure IMC and pure DDMC are similar.

The DDMC scaled computation times for the strong redshift coupling test are tabulated
below.
\begin{table}[H]
\caption{Strong Group Coupling Computation Times}
\centering
\resizebox{40 mm}{!}{
\begin{tabular}{|l|r|}
\hline
Method & Scaled Time \\
\hline
DDMC & 1 \\
\hline
HMC & 44.89\\
\hline
IMC & 110.39\\
\hline
\end{tabular}
}
\end{table}

\begin{figure}
\subfloat[]{\includegraphics[height=70mm]{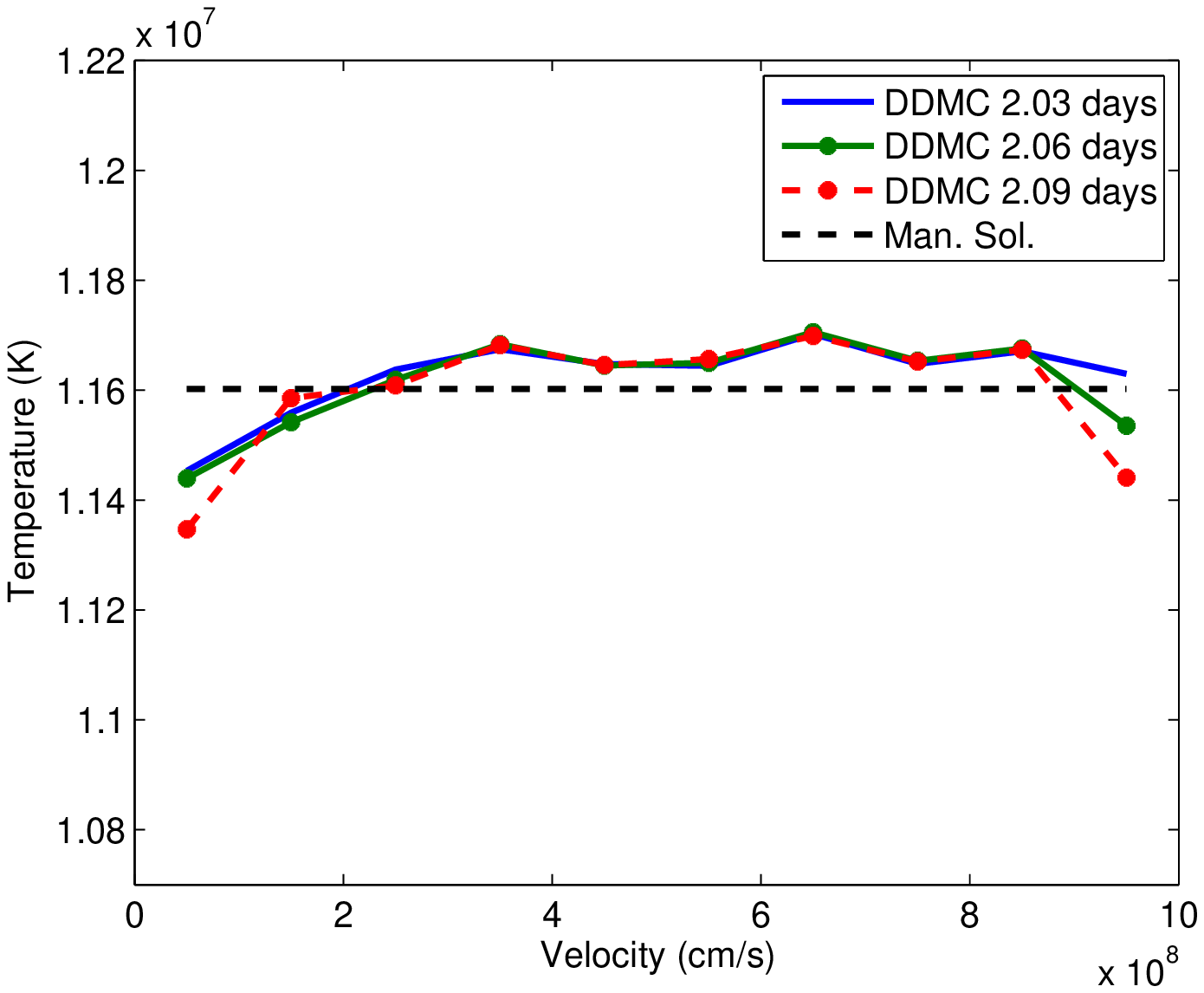}\label{fg11}}\\
\subfloat[]{\includegraphics[height=70mm]{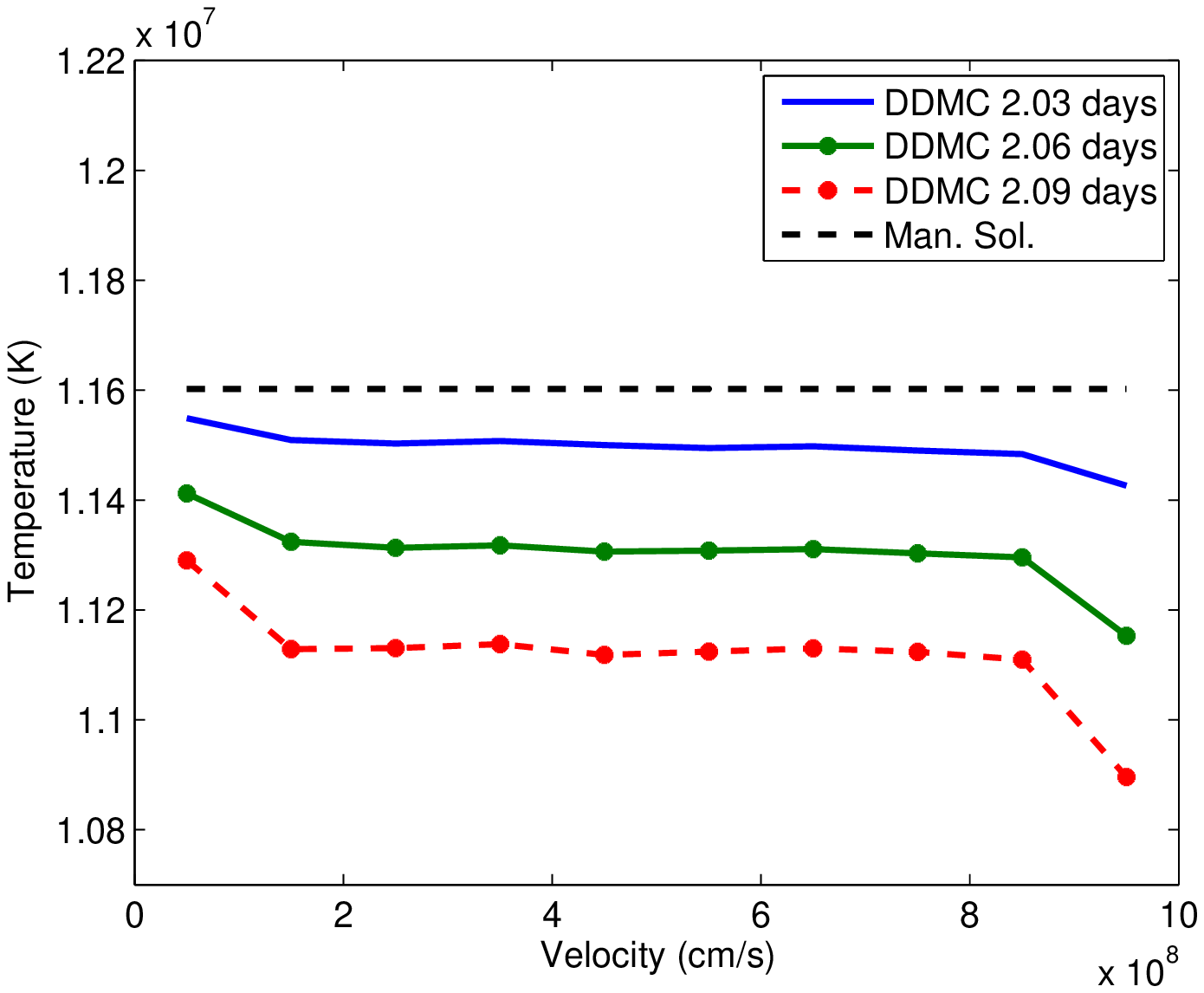}\label{fg12}}
\caption{
Manufactured (dashed) and DDMC (solid, dot-solid, and dot-dashed) material temperature
at three different times plotted over the velocity grid.  The MC profiles in Fig.~\ref{fg11}
are generated by implementing the manufactured source described in Section~\ref{sec:Manu}
with strong between-group redshift coupling.  It is evident that a steady state is
maintained at the inner regions of the domain.  The MC profiles is Fig.~\ref{fg12} are
generated in nearly the same manner but leave the lower wavelength ($g=1$) group sourceless.
In Fig.~\ref{fg12}, the non-equilibrium in temperature is a result of the higher wavelength
($g=2$) group not being able to produce a steady state temperature without an additional
source of Doppler shifted radiation (from $g=1$).  The purely elastic scattering $g=1$
group thus has an indirect yet important effect on the temperature for the strong Doppler
coupling described.
}
\end{figure}

We ensure the pure scattering group, $g=1$, indirectly sustains the temperature
profile's steady state by removing the radiation in that group.  The profile should
steadily drop relative to the solution that includes Doppler shifting.  Evidence for
this effect can be found in Figs.~\ref{fg11} and~\ref{fg12}.

\newpage
\subsection{Spherical Heaviside Source Tests}
\label{sec:Heav}

\begin{figure*}
\subfloat[]{\includegraphics[height=70mm]{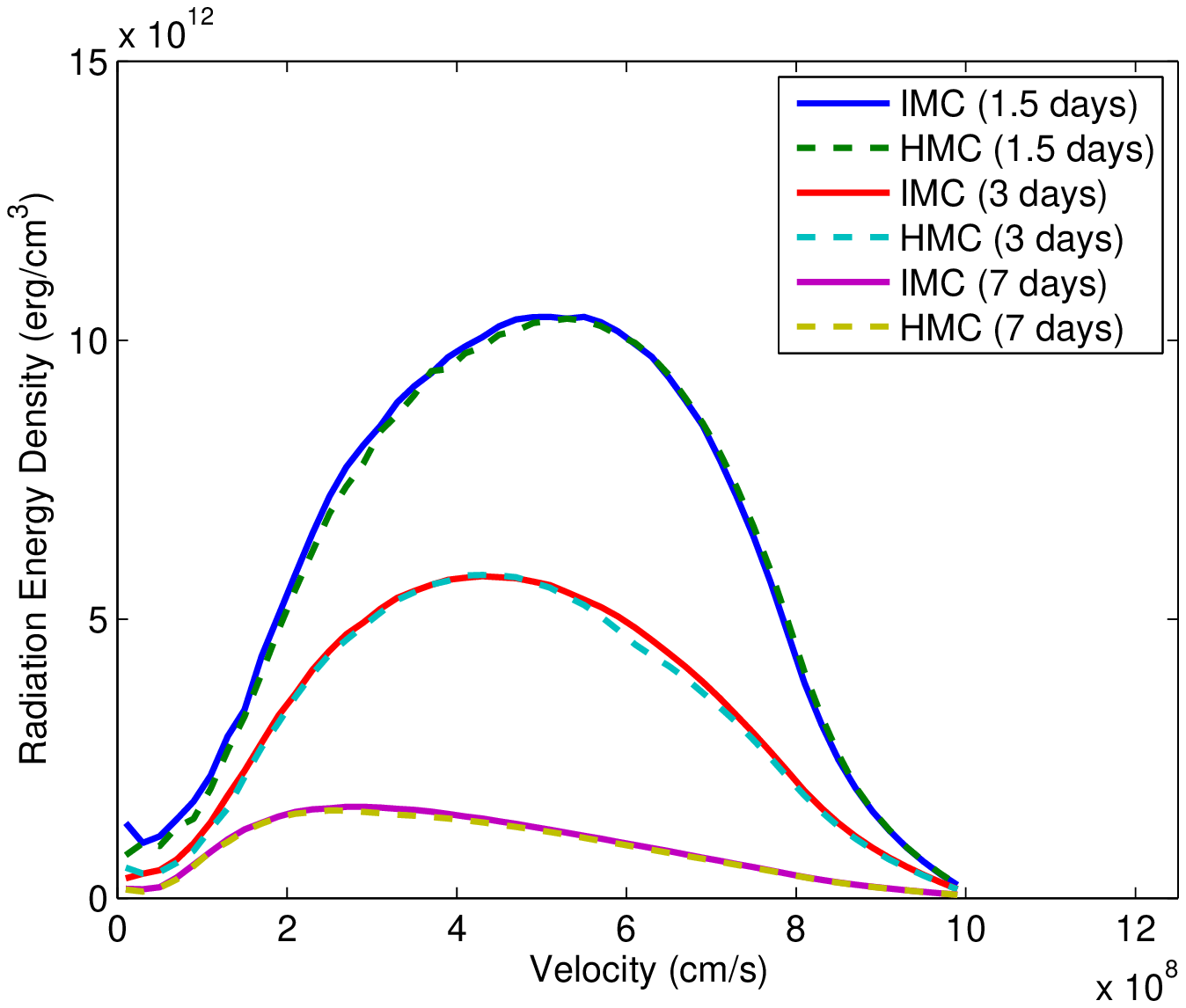}\label{fg13}}
\subfloat[]{\includegraphics[height=70mm]{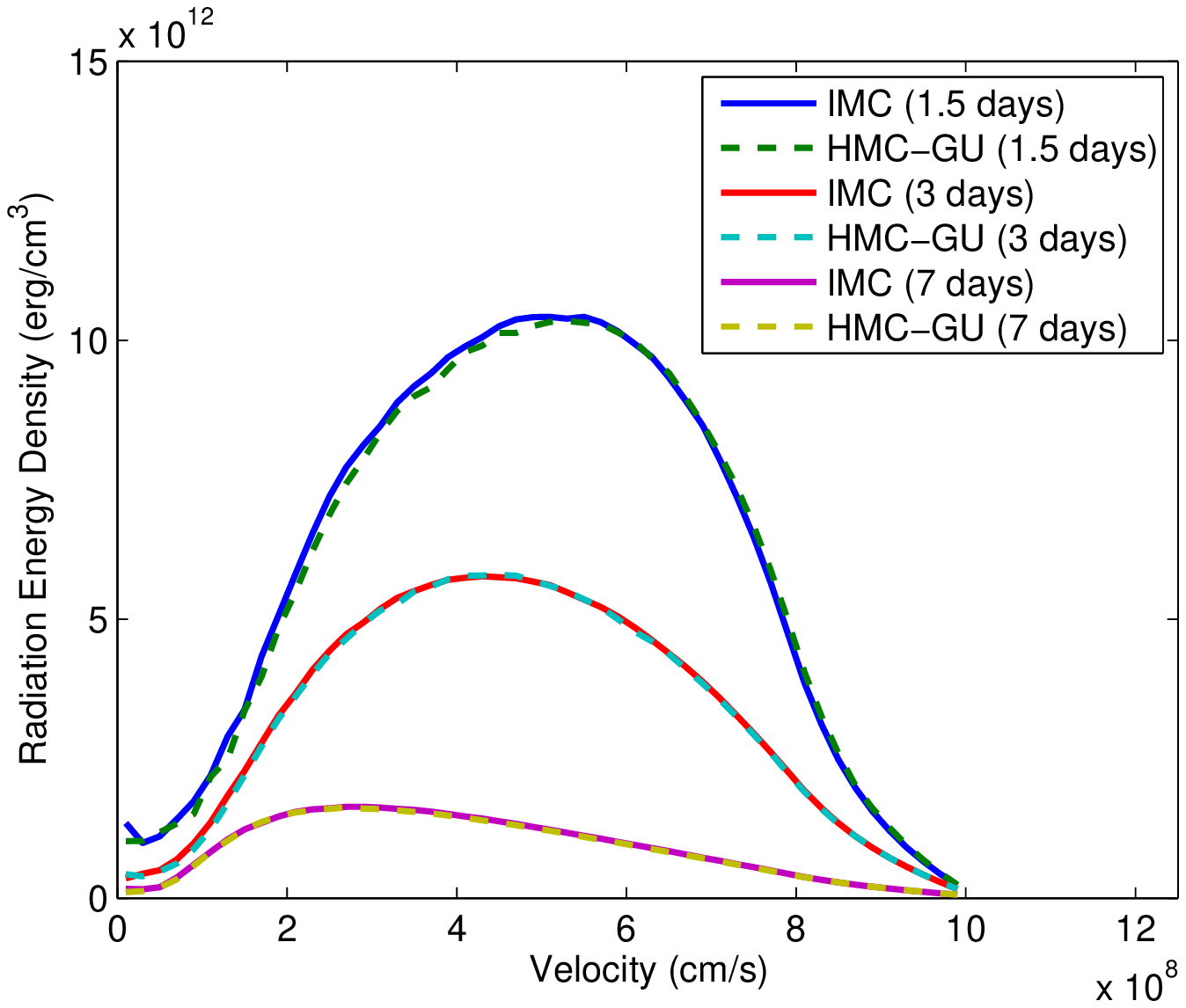}\label{fg15}}
\caption{
Radiation energy density for pure IMC (solid lines) and IMC-DDMC (dashed lines)
at three times for an expanding domain with a $4\times10^{24}/(t_{n}+t_{\min})^{3}$ erg/cm$^{3}$/s spherical
Heaviside source from the expansion center out to the location of the fluid moving at $8\times10^{8}$ cm/s.  The
opacity is defined over 10 logarithmic wavelength groups with a magnitude of $0.13\rho$ cm$^{-1}$ in odd groups and
$0.13\times 10^{-4}\rho$ cm$^{-1}$ in even ($\rho$ is density in g/cm$^{3}$).  Because equal mass is attributed to
each of the fifty fluid cells, the density is not radially uniform.  At day 1.5, most particles are propagating with DDMC.
By day 7, most particles are propagating with IMC.  For this problem, $\tau_{\text{DDMC}}=3$ and IMC-DDMC is faster than
IMC by a factor of 3.36.
In Fig.~\ref{fg13}, the standard IMC-DDMC boundary condition is used; at 3 and 7 days the boundary discrepancy has propagated
to about $6\times10^{8}$ cm/s and $4\times10^{8}$ cm/s, respectively.  In Fig.~\ref{fg15}, we have implemented the
fit, Eq.~\eqref{eq79} with $C_{1}=0.55$ and $C_{2}=1.25$, of the amplification factor, and find for this problem, with
the IMC-DDMC threshold at 3 mean free paths, that the method boundary induced error has essentially been removed.
}
\end{figure*}

\begin{figure}
\includegraphics[height=70mm]{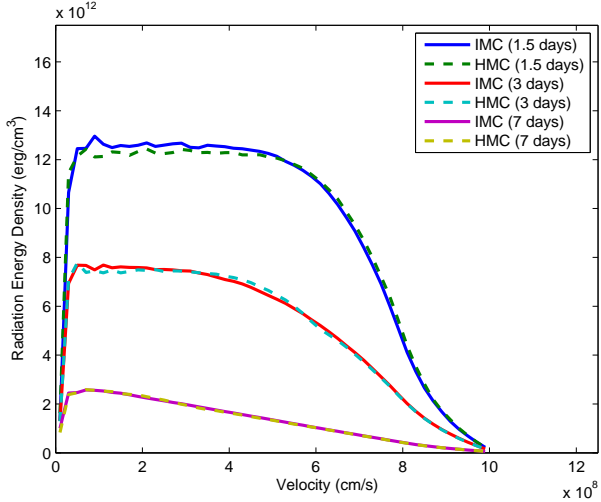}
\caption{
Radiation energy density for pure IMC (solid lines) and IMC-DDMC (dashed lines)
at three times for an expanding domain with a $4\times10^{24}/(t_{n}+t_{\min})^{3}$ erg/cm$^{3}$/s spherical Heaviside
source from the expansion center out to the location of the fluid moving at $8\times10^{8}$ cm/s.  The opacity is
defined over 10 logarithmic wavelength groups with a magnitude of $0.13\rho$ cm$^{-1}$ in odd groups and
$0.13\times 10^{-7}\rho$ cm$^{-1}$ in even ($\rho$ is density in g/cm$^{3}$).  Because equal mass is attributed to
each of the fifty fluid cells, the density is not radially uniform.  At day 1.5, most particles are propagating with DDMC.
By day 7, most particles are propagating with IMC.  For this problem, IMC-DDMC is faster than IMC by a factor of 4.59.
}
\label{fg14}
\end{figure}

\begin{figure}
\includegraphics[height=70mm]{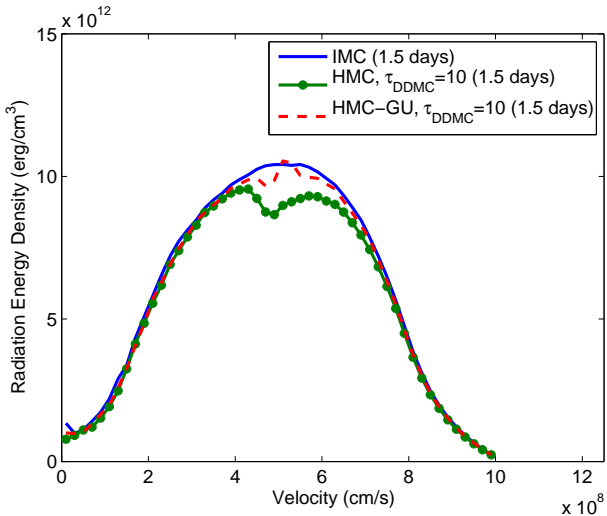}
\caption{
Radiation energy density for pure IMC (solid line), IMC-DDMC without a
$G_{U}$ factor (dotted solid line), and IMC-DDMC with Eq.~\eqref{eq79}
(dashed line) at 1.5 days after initial time with a $4\times10^{24}/(t_{n}+t_{\min})^{3}$
erg/cm$^{3}$/s spherical Heaviside source from the expansion center out to the location of
the fluid moving at $8\times10^{8}$ cm/s.  The opacity is defined over 10 logarithmic
wavelength groups with a magnitude of $0.13\rho$ cm$^{-1}$ in odd groups and
$0.13\times 10^{-4}\rho$ cm$^{-1}$ in even ($\rho$ is density in g/cm$^{3}$).  The
introduction of the $G_{U}$ factor into IMC-DDMC has improved
agreement with pure IMC for $\tau_{\text{DDMC}}=10$, but error persists.
Additionally, some Monte Carlo noise is added due to the large range of $G_{U}(\mu)$
modifying particle weights.  To avoid this issues for the method described, we
recommend a $\tau_{\text{DDMC}}$ threshold between 2 and 5.
}
\label{fg16}
\end{figure}

Finally, we construct a multigroup outflow problem that may be used to test both code efficiency and
quality of the operator split treatment of grid motion in IMC-DDMC.  The problem consists of a
Heaviside spherical source out to $0.8U_{\max}$ of strength 4$\times 10^{24}/(t_{n}+t_{\min})^{3}$
ergs/cm$^{3}$/s in a purely absorbing fluid.  The opacity is over 10 groups logarithmically spaced in
wavelength from 1.238$\times10^{-9}$ cm to 1.238$\times10^{-3}$ cm.  As in the manufactured solution,
we alternate small and large opacities across the wavelength grouping where odd groups have the larger
opacity.  Again, the speed $U_{\max}=10^{9}$ cm/s and the total mass $M=10^{33}$ g.  Instead of equal density
in each cell, we attribute $M/J$ mass to each cell for $J$ fluid cells.  Uniformly partitioning the mass 
creates a density gradient that couples into the transport process through the macroscopic opacity.  All 
simulations in this section use 50 velocity cells from 0 cm/s to $U_{\max}$, 192 time steps from 2 days to 11 days,
and 100,000 source particles per time step.  The heat capacity is adopted from~\cite{pinto2000};
$C_{v}\approx 2\times10^{7}\rho$ ergs/cm$^{3}$/K.  For the calculations discussed, we use DDMC for cell $j$
when $\Delta U_{j}\Delta t_{n}\sigma_{g}\geq\tau_{\text{DDMC}}$ and IMC otherwise.
Following~\cite{abdikamalov2012},
we have labeled the threshold mean free path number between IMC and DDMC as $\tau_{\text{DDMC}}$.
For the first two tests discussed, depicted in Figs.~\ref{fg13} and~\ref{fg14}, $\tau_{\text{DDMC}}=3$.
For the tests of the new boundary condition, depicted in Figs.~\ref{fg15} and~\ref{fg16}, $\tau_{\text{DDMC}}=3$
for Fig.~\ref{fg15} and $\tau_{\text{DDMC}}=10$ for Fig.~\ref{fg16}.
As a consequence of the problem's structure, the density
gradient creates a ``method front'' for IMC-DDMC where an outer shell of IMC moves inward over the grid.
The method front is
heterogeneous in group space, meaning the even groups are converted to IMC sooner (or are already IMC) over
the specified problem duration.

The discrepancies between the IMC and IMC-DDMC profiles in Figs.~\ref{fg13} and~\ref{fg14} arise in part
from implementing Eq.~\eqref{eq32} instead of Eq.~\eqref{eqAB}.  For mean free path thresholds on the order of 2
to 3 per cell, we find that these errors are systematic yet generally minimal.  The discrepancy can be made much worse
by implementing a more conservative mean free path threshold of about 10 for this problem set.  Higher mean free
path thresholds require IMC particles emitted from a DDMC spatial surface to propagate through an optically
thicker sub-cell environment.  Of course this discrete interface is not present for pure IMC; in other words there
is not a significant source of IMC radiation originating from one surface in pure IMC since the method interface
is not present.  Complementarily, the DDMC field is not as sourced by IMC radiation for a high mean free path
threshold.  This is discernible from the form of $P_{b(j,j')}(\mu)$.  We demonstrate that incorporating an
approximation of the new factor, $G_{U}$, from Eq.~\eqref{eqAB} indeed appears to mitigate the over-redshift near
the method front in Figs.~\ref{fg15} and~\ref{fg16}.  However, we recommend a mean free path threshold in
the range of 2 to 5 mean free paths.  Arguments presented by~\cite{pinto2000} indicate that diffusion theory
remains valid on large outflow time scales if radiation momentum does not greatly affect fluid momentum.

In the first case tested, we use
\begin{equation}
\label{eq77}
\sigma_{g}=\begin{cases}
0.13\rho \;\;,\;\;g=2k-1\\
0.13\times 10^{-4}\rho \;\;,\;\;g=2k \;\;.
\end{cases}
\end{equation}
Results for Eq.~\eqref{eq77} are plotted in Fig.~\ref{fg13}.  We find IMC-DDMC is faster than IMC
by a factor of 3.36.

In the second case tested, we use
\begin{equation}
\label{eq78}
\sigma_{g}=\begin{cases}
0.13\rho \;\;,\;\;g=2k-1\\
0.13\times 10^{-7}\rho \;\;,\;\;g=2k \;\;.
\end{cases}
\end{equation}
Results for Eq.~\eqref{eq78} are plotted in Fig.~\ref{fg14}.  We find IMC-DDMC is faster than IMC
by a factor of 4.59.  We have improved the performance of IMC-DDMC relative to IMC by increasing
the disparity in adjacent group opacities.  If instead the $g=2k-1$ opacities are increased,
IMC-DDMC provides even further improvement in the diffusive groups of the spectrum.  In both
simulations, IMC-DDMC transitions to pure IMC over the 9 day period.

We now describe a possible implementation of Eq.~\eqref{eqAB}.  There are two readily
discernible means of ascribing a MC interpretation to the $G_{U}(\mu)$ factor.  In one, the
probability that an IMC particle transmits into DDMC when incident on a diffusive region
may be taken as $P(\mu)\sim (1+3\mu/2)G_{U}(\mu)$.  However, this would make $P(\mu)>1$
for some $\mu$ regardless of cell material properties.  To constrain $P(\mu)\leq 1$, some
range of angular projections, $\mu\in[0,\varepsilon]$, where $\varepsilon<1$, would have to
have $G_{U}$ modified.
As an alternative, the probability of IMC to DDMC transmission may be maintained as its original
form $P(\mu)\sim 1+3\mu/2$ and the particle weight must then be multiplied by $G_{U}(\mu)$.
Since $G_{U}(\mu)$ is of O($1/\mu$), the energy current across the IMC-DDMC
interfaces is bounded.  Physically, it is supposed that this implies the importance of IMC particles
interacting with a surface at a DDMC region to the energy balance in the adjacent cells must still
be bounded at small values of $\mu$.  In our tests, for an inner DDMC region adjacent to an outer IMC
region at a boundary with speed $U$,
\begin{equation}
\label{eq79}
G_{U}(\mu)=1+2\frac{U}{c}\left(\frac{C_{1}}{\mu}-C_{2}\mu\right) \;\;,
\end{equation}
where $C_{1},C_{2}>0$ are constants.  In passing, we note that a angularly uniform, heuristic
$G_{U}$ may be calibrated from simulation.  From phenomenological considerations, it is found that
a good form of $G_{U}$ is
\begin{equation}
\label{eq80}
\bar{G}_{U}=1+2\min(0.055\tau_{\text{DDMC}},1)\frac{U}{c}
\end{equation}
at least for a range of $\tau_{\text{DDMC}}\in [3,10]$.
Applying Eq.~\eqref{eq79} with $C_{1}=0.55$ and $C_{2}=1.25$
for $\tau_{\text{DDMC}}=\Delta U_{j}\Delta t_{n}\sigma_{g}=3$ and Eq.~\eqref{eq77}, we see a
small improvement in Fig.~\ref{fg15}.

To further demonstrate the potential utility of
Eq.~\eqref{eq79}, we apply the $G_{U}$ factor to a problem where the discrepancy is made very
large by setting $\tau_{\text{DDMC}}=10$.
The fitting constants for this test are $C_{1}=0.6$ and $C_{2}=1.25$.
We observe a significant improvement at all times including when the discrepancy is very large
at 1.5 days, as plotted in Fig.~\ref{fg16}.  However, with this improvement comes some
additional MC noise due to the weight modifications having a large range of $G_{U}$ and insufficient
sampling for $\mu\rightarrow 0$.  Since the IMC portion of the simulation is in the lab frame, the
minimal comoving projection into the DDMC interface cell is $U/c$.  But physically, the comoving
projection lower bound is 0.  For sampling
comoving directions with $\mu<U/c$, the new boundary condition is applied to particles with $\mu<U/c$
that are advected onto DDMC surfaces through the velocity position shift algorithm
delineated in Section~\ref{subsec:MFIMCDDMC}.  We find that these samplings are not important for
$\tau_{\text{DDMC}}=3$ but are important for $\tau_{\text{DDMC}}=10$ or greater.  At larger
$\tau_{\text{DDMC}}$, IMC particles that ``graze'' the DDMC surface are important due to the increased
level of angular isotropy near the IMC-DDMC boundary.  We note that the proper implementation
of the weight modifying factor, $G_{U}$, is an open research question.  Additionally, our choice
of asymptotic scalings and analysis in Section~\ref{subsec:MBLA} is approximate and tailored for
homologous outflow.

We caution that what may be thought of as a conservative selection of a mean
free path threshold between IMC and DDMC may lead to the types of errors
depicted in Fig.~\ref{fg16}.

\section{Conclusions and Future Work}
\label{sec:Conc}

We have described an approach to multifrequency 
IMC-DDMC on a ``velocity grid'' that is semi-implicit, accelerated by diffusion theory, and relativistic to
first order.  Additionally, we have provided an algorithm for treating the operator split motion of Lagrangian
grid boundaries that is simple to integrate into any IMC-DDMC scheme possessing hydrodynamic effects.
This treatment of radiation transport is a viable candidate for simulating the post-explosion phase of
thermonuclear supernovae.

In Sections~\ref{sec:Static} and~\ref{sec:Manu}, we have provided a simple generalization of McClarren's
analytic P$_{1}$ solutions~\citep{mcclarren2008} for static material multifrequency verification and a
manufactured solution for multigroup outflow verification.  We find that {\tt SuperNu} produces good agreement with
both analytic solutions.

In Section~\ref{sec:Heav}, we perform spherical Heaviside source simulations with high-disparity grouped
opacities in the presence of a fluid density gradient.  This density gradient induces a ``method front'' in
IMC-DDMC where an inward moving region of pure IMC starting at the outermost fluid cell replaces DDMC in the
optically thick groups.  For our tests, IMC-DDMC produces good agreement with pure IMC at all method front
locations over the velocity grid; this indicates that the Lagrangian grid boundary algorithm and the
particular choice of mean-free-path based coupling between IMC and DDMC are functioning properly.

We have discovered that when the fluid velocity is semi-relativistic, an important correction
term is necessary at IMC-DDMC boundaries.  In our findings, we have seen that errors over 10\% in radiation
energy density may manifest in IMC-DDMC relative to pure IMC for reasonable input parameters if the standard IMC-DDMC
boundary condition is used.  The corrective term generally reduces these boundary errors and increases the
viable range of IMC-DDMC mean-free-path thresholds.
However, we note that the best results observed are for thresholds on the order of 2 to 5 mean free paths even
with the new boundary condition.
Despite the singularity in the
new factor, the influence of all particles on the energy balance in each cell
is finite by virtue of the analysis performed in Section~\ref{subsec:MBLA}.  In Section~\ref{sec:Heav},
we have shown that these discrepancies may occur in astrophysical problems.
Also in Section~\ref{sec:Heav}, we have used the modified IMC-DDMC boundary condition to indeed improve
agreement between IMC-DDMC and pure IMC for different values of mean free path threshold, $\tau_{\text{DDMC}}$.
We note that the theory and implementation of the corrective boundary factor presented here requires further
exploration.  These Heaviside tests additionally demonstrate that IMC-DDMC
performs much better than pure IMC in terms of accuracy and speed when there are large disparities between
the magnitudes of opacities in adjacent groups, which is the primary motivation of this work.

We plan to incorporate nonuniform frequency or
wavelength groups across spatial cells~\citep{densmore2012}.
In order to do so, fully general phase space leakage graphs must be implemented.
These graphs (qualitatively depicted in Fig.~\ref{fg1}) along with group lumping
(Eqs.~\eqref{eq43}-\eqref{eq45}) may help improve efficiency for transport problems
with many groups and ill-behaved spectral properties.  Additionally, we plan to further investigate alternative
implementations of the theory presented in Section~\ref{subsec:MBLA}.  The initial target application for the
one dimensional, spherically symmetric code is Nomoto's W7 model~\citep{nomoto1984}.  We subsequently intend to
implement the IMC-DDMC method in multiple dimensions.  On the basis of the preliminary evidence accrued, we expect
to achieve a diffusion-accelerated, implicit treatment of radiation in resolved SNe simulations that is robust
and scalable.

\section*{Acknowledgements}
We would like to thank Jeffrey Densmore, and Allan Wollaber for their
very helpful insight, exchanges and sources.
We especially thank our referee, Ernazar Abdikamalov, for valuable discussions and a thorough review
that improved this paper.
This work was supported in part by the University of Chicago and the National Science Foundation
under grant AST-0909132.
SMC is supported by NASA through Hubble Fellowship grant No. 51286.01 awarded by the Space
Telescope Science Institute, which is operated by the Association of Universities for Research in Astronomy,
Inc., for NASA, under contract NAS 5-26555.

\nocite{*}
\bibliography{Bibliography}

\end{document}